# Model Based System Assurance Using the Structured Assurance Case Metamodel

Ran Wei[a,1,*], Tim P. Kelly[a,*], Xiaotian Dai[a,*], Shuai Zhao[a,*], Richard Hawkins[a]

[a]*Department of Computer Science, University of York, York, YO10 5GH, UK*

## Abstract

Assurance cases are used to demonstrate confidence in system properties of interest (e.g. safety and/or security). A number of system assurance approaches are adopted by industries in the safety-critical domain. However, the task of constructing assurance cases remains a manual, trivial and informal process. The Structured Assurance Case Metamodel (SACM) is a standard specified by the Object Management Group (OMG). SACM provides a richer set of features than existing system assurance languages/approaches. SACM provides a foundation for model-based system assurance, which has great potentials in growing technology domains such as Open Adaptive Systems. However, the intended usage of SACM has not been sufficiently explained. In addition, there has been no support to interoperate between existing assurance case (models) and SACM models.

In this article, we explain the intended usage of SACM based on our involvement in the OMG specification process of SACM. In addition, to promote a model-based approach, we provide SACM compliant metamodels for existing system assurance approaches (the Goal Structuring Notation and Claims-Arguments-Evidence), and the transformations from these models to SACM. We also briefly discuss the tool support for model-based system assurance which helps practitioners to make the transition from existing system assurance approaches to model-based system assurance using SACM.

*Corresponding author

*Email addresses:* ran.wei@york.ac.uk (Ran Wei), tim.kelly@york.ac.uk (Tim P. Kelly), xiaotian.dai@york.ac.uk (Xiaotian Dai), shuai.zhao@york.ac.uk (Shuai Zhao)

## 1. Introduction

Systems and services used to perform critical functions require justifications that they exhibit necessary properties (i.e. safety and/or security). ***Assurance case***s provide an explicit means for justifying and assessing confidence in these critical properties. In certain industries, typically in the safety-critical domain, it is a regulatory requirement that an assurance case is developed and reviewed as part of the certification process [11]. An assurance case is a document that facilitates information exchange between various system stakeholders (e.g. between operator and regulator), where the knowledge related to the safety and/or security of the system is communicated in a clear and defend-able way [11].

Assurance cases are typically represented either textually - using natural languages; or graphically - using structured graphical notations such as the Goal Structuring Notation (GSN) [18] or Claims-Arguments-Evidence (CAE) [4]. Graphical notations have gained popularity due to their abilities to express clear and well structured argumentations. A number of tools exist which implement GSN and CAE to produce safety cases, which are surveyed in [24]. Some tools adopt Model-Driven Engineering (MDE) to produce models that conform to their own versions of GSN/CAE metamodels [8; 26; 27; 21; 3].

To improve standardisation and interoperability, the Object Management Group (OMG) specified and issued the Structured Assurance Case Metamodel (SACM). SACM is developed by the specifiers of existing system assurance approaches (e.g. GSN and CAE), based on the collective knowledge and experiences of safety/security practitioners. Comparing to existing assurance case approaches, SACM provides additional features such as fine-grained modularity, controlled vocabulary, and argument-evidence traceability. Thus, SACM is more powerful in terms of expressiveness. However, a detailed explanation of how to use SCAM is not provided in the OMG specification. In addition, the relationships between existing assurance case approaches (i.e. GSN and CAE) and SACM have not been sufficiently discussed. These bring challenges to the adoption of SACM due to the complexity of SACM and the sophistication of its intended usage.

Model-based system assurance has attracted a significant amount of interests in recent years due to the benefits provided by MDE such as automation and consistency. Model-based system assurance is particularly important for concepts such as Open Adaptive Systems (OAS), where open (safety/security critical) systems connect to each other, and adapt to changing contexts at runtime.

As the principal contributors of SACM and the originators of GSN, in this paper, we provide a detailed explanation of SACM and discuss its relationship with existing system assurance approaches (i.e. GSN and CAE). The contributions of this paper are:

- A definitive exposition of SACM version 2.0;
- An explanation on how to create assurance case models using SACM;

- A presentation of the GSN and CAE metamodels that are compliant with SACM;
- Comprehensive mappings from GSN/CAE to SACM.

This paper is organised as follows. In Section 2 and Section 3, we provide the background and the motivation of our work. In Section 4 we provide detailed discussions about the facilities provided by SACM. In Section 5 we provide examples to illustrate the semantics of the elements provided in SACM, and how to use SACM to construct argumentation patterns and to integrate assurance cases. In Section 6, we discuss the relationship between existing notations and SACM. We provide SACM compliant metamodels for GSN and CAE and their mappings to SACM. In Section 7, we briefly discuss tool support for model-based system assurance. Finally, a conclusion of this paper is given in Section 8.

## 2. Background and Motivation

### 2.1. Safety Cases

The concept of assurance cases has been well established in the safety-related domains, where the term ***safety case*** is normally used. For many industries, the development, review and acceptance of a safety case form a key element of regulatory processes. This includes nuclear, defence, aviation and railway industries [14]. Safety cases are defined in [18] as follows: ***A safety case should communicate a clear, comprehensible and defensible argument that a system is acceptably safe to operate in a particular context.***

Historically, safety arguments were typically communicated in safety cases through free text. However, there are problems experienced when text is the only medium available for expressing complex arguments. One problem of using free text is that the language used in the text can be unclear, ambiguous and poorly structured. There is no guarantee that system engineers would produce safety cases with a clear and well-structured language. Also, the capability of expressing cross-references with free text is very limited, multiple cross-references can also disrupt the flow of the main argument. Most importantly, the problem with using free text is in ensuring that all stakeholders involved share the same understanding of the argument to develop, agree and maintain the safety arguments within the safety case [18].

To overcome the problems of expressing safety arguments in free text, graphical argumentation notations were developed. Graphical argumentation notations are capable of explicitly representing the elements that form a safety argument (i.e. requirements, claims, evidence and context), and the relationships between these elements (i.e. how individual requirements are supported by specific claims, how claims are supported by evidence, and the assumed context that is defined for the argument). Amongst the graphical notations, the ***Goal Structuring Notation*** (GSN) [18] has been widely accepted and adopted [6]. The key benefit experienced by companies/organisations adopting GSN is that it improves the comprehension of the safety argument among all key project stakeholders (e.g. system developers, safety engineers, independent assessors and certification authorities), therefore improving the quality of the debate and discussion amongst the stakeholders and reducing the time taken to reach agreements on the argument approaches being adopted.

Another popular graphical argumentation notation is ***Claims-Arguments-Evidence*** (CAE) [4]. CAE views assurance cases as a set of ***Claim***s supported by ***Argument*** s, which in turn rely on ***Evidence***. Compared to CAE, GSN provides more granular decomposition of satety arguments, and supports additional features such as modularity and argument patterns [17]. In this paper, we focus on GSN and its relationship to SACM, since we are the principal contributors to the standardisation of both GSN and SACM.

A number of graphical assurance case tools have been developed due to the popularity of graphical argumentation notations. A recent study [24] has looked into and compared assurance case tools that have been developed in the past twenty years (where 32 of them support GSN). The majority of the tools do not support model-based approach (i.e. creating model-based graphical assurance cases).

### 2.2. Safety Cases and Model Driven Engineering

Model-Driven Engineering (MDE) is a contemporary software engineering approach. In MDE, ***model***s are first class artefacts, therefore ***driving*** the development. There are two important aspects in MDE: ***Domain Specific Modelling*** and ***Model Management***. ***Domain Specific Modelling*** enables domain experts to capture the concepts in their systems in the form of ***metamodel*** s, which are then used to create models of their systems (that conform to the defined ***metamodel***s). ***Model Management*** enables a variety of operations to be performed on models in an automated manner, which include, but not limited to: Model Validation, Model-to-Model Transformation, Model-to-Text Transformation and Model Merging. MDE has been proven to improve consistency and productivity significantly due to the automation provided by model management operations [15; 16].

Whilst graphical assurance cases are valuable in communicating the safety and/or security properties of the system, MDE is beneficial to system assurance to enable higher level model management operations to be performed. Provided that assurance cases are constructed as models, model validation can be used to check well-formedness of assurance cases (e.g. in GSN, a ***Strategy*** cannot be directly supported by a ***Solution***), model-to-text transformation can be used to generate assurance case reports, and model merging can be used to bind assurance cases. A number of assurance case tools adopt MDE, such as AdvoCATE [8], D-Case Editor [26], ASCE [27], Astah GSN [21], and CertWare [3]. However, a common problem with these tools is that they define their own GSN metamodels. This is due to the fact that there has not been a standardised GSN metamodel[1]. Thus, there may be interoperability problems when one wishes to

[1]Apart from the mappings from GSN to SACM provided on the GSN website – http:

import GSN models created by other tools into his/her own tool. Although the tools mentioned above all claim to support SACM, the support is for SACM version 1.0 (released in June, 2015), which was replaced by SACM version 2.0 (released in March, 2018) with significant changes to the specification[2]. Since SACM is not sufficiently explained (no work has been done in this aspect), there may be cognitive gaps between different tool developers, so that exported SACM models from the tools may differ.

### *2.3. SACM and Runtime System Assurance*

Model based system assurance is particularly necessary in assuring emerging system concepts such as Open Adaptive System - OAS (e.g. Cyber-Physical Systems - CPS). Industry sees huge economic potential in such systems - particularly due to their open (OASs connect to each other at runtime) and adaptive (OASs adapt to changing contexts) nature, which enables new types of promising applications in domains such as automotive[3], health care, and home automation [32]. Since the majority of application domains of OAS are safety-critical, it is imperative to assure the safety and/or security of such systems. However, assuring OASs is typically difficult. Due to their open and adaptive nature, it is almost impossible to sufficiently anticipate the concrete system structure, the system' s capabilities and the environmental context at design time. [31].

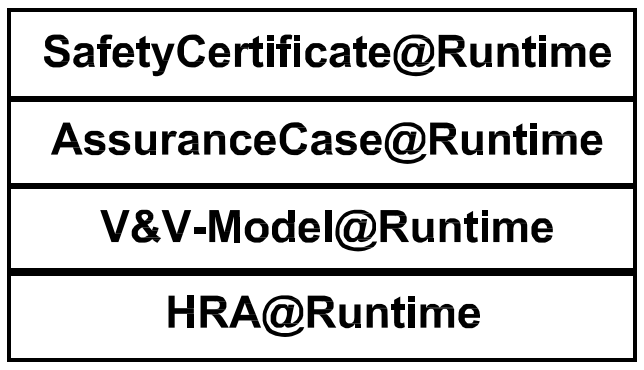

Figure 1: Models at Runtime on Different Abstraction Levels.

Therefore, existing design time system assurance activities are insufficient to enable dynamic system assurance for OAS at runtime. In [31], the authors identify the importance of system assurance at runtime for OAS and propose the idea of ***Models@Runtime*** on different abstraction levels, as shown in Figure 1. On the top abstraction level is the SafetyCertificate@Runtime, which is a contract-like model that contains information with regard to services and guarantees provided/demanded. With SafetyCertificate@Runtime, systems can determine if they can interact with each other. However, guarantees are not enough to infer assurance of systems at runtime, the process of argumentation

//www.goalstructuringnotation.info/

[2]ASCE claims that it supports SACM 2.0 but the mappings they provide in https://www.adelard.com/asce/choosing-asce/standardisation.html are still mapping to old SACM elements

[3]Autonomous cars being a typical example

that leads to the claims about the safety/security of a system is needed. Hence, AssuranceCases@Runtime is necessary so that one OAS can inspect the assurance case of other OASs at runtime to conclude if they are safe to interact with. For autonomous systems, some safety claims can only be instantiated at runtime due to the uncertainty introduced by Artificial Intelligence. Hence, a runtime safety case is needed for autonomous systems to infer the safety/security of themselves (with the help of assurance case managing facilities). When system adapt, the evidence for system assurance may become invalid. Therefore, Verification and Validation (V&V) models at runtime may help to reinstate evidence that is used by assurance case models. In very rare cases, system requirements may change at runtime. Creating Hazard and Risk Analysis (HRA) models at runtime helps identify new hazards and risks introduced based on the change of requirements. As explained in [33], V&V and HRA models at runtime are not realistic, because they are effort intensive tasks which require human intervention and cannot be automated. On the other hand, SafetyCertificate@Runtime is not sufficient to determine if a system is safe/secure. The ideal balance of system assurance at runtime is on the AssuranceCase@Runtime, where a system can use assurance cases to relate to other models at runtime to determine the safety/security of itself, or for a system to determine the safety/security of other systems. Thus, there is a need to shift design time assurance case documents to runtime assurance case models, in order to assure open adaptive systems at runtime.

Comparing to GSN and CAE, SACM provides an ideal basis that could underpin AssuranceCase@Runtime in this context, due to the features mentioned later in this section. In [8], the authors motivate the need for automation in assurance case. They point out that assurance case models should link their evidence in order to perform automated reasoning on the validity of assurance cases. However, on modelling language level, GSN and CAE are not capable for this task. In a first step towards ***runtime*** system assurance, SACM is used in the DEIS project [32] as a backbone for its Open Dependability Exchange metamodel (ODE), which is used to assure the dependability of Cyber-Physical Systems at runtime, as illustrated in Figure 2. System assurance at runtime is achieved by having an assurance case model which links to ODE models (as evidence) to form the argumentation for system assurance[4]. At runtime, ODE models are changed based on the sensor data. There is a certification algorithm which periodically evaluates the assurance case. Should any ODE models change at runtime, re-certification of the assurance case is performed. The system at runtime executes the assurance case and determines if it is safe/secure to operate. SACM provides a solid foundation for practitioners to argue about the safety and/or security of a wide range of systems, and to refer to the evidence of their arguments as models, both at design time and runtime.

[4]This feature is supported by SACM but not GSN/CAE, which is discussed later in the article.

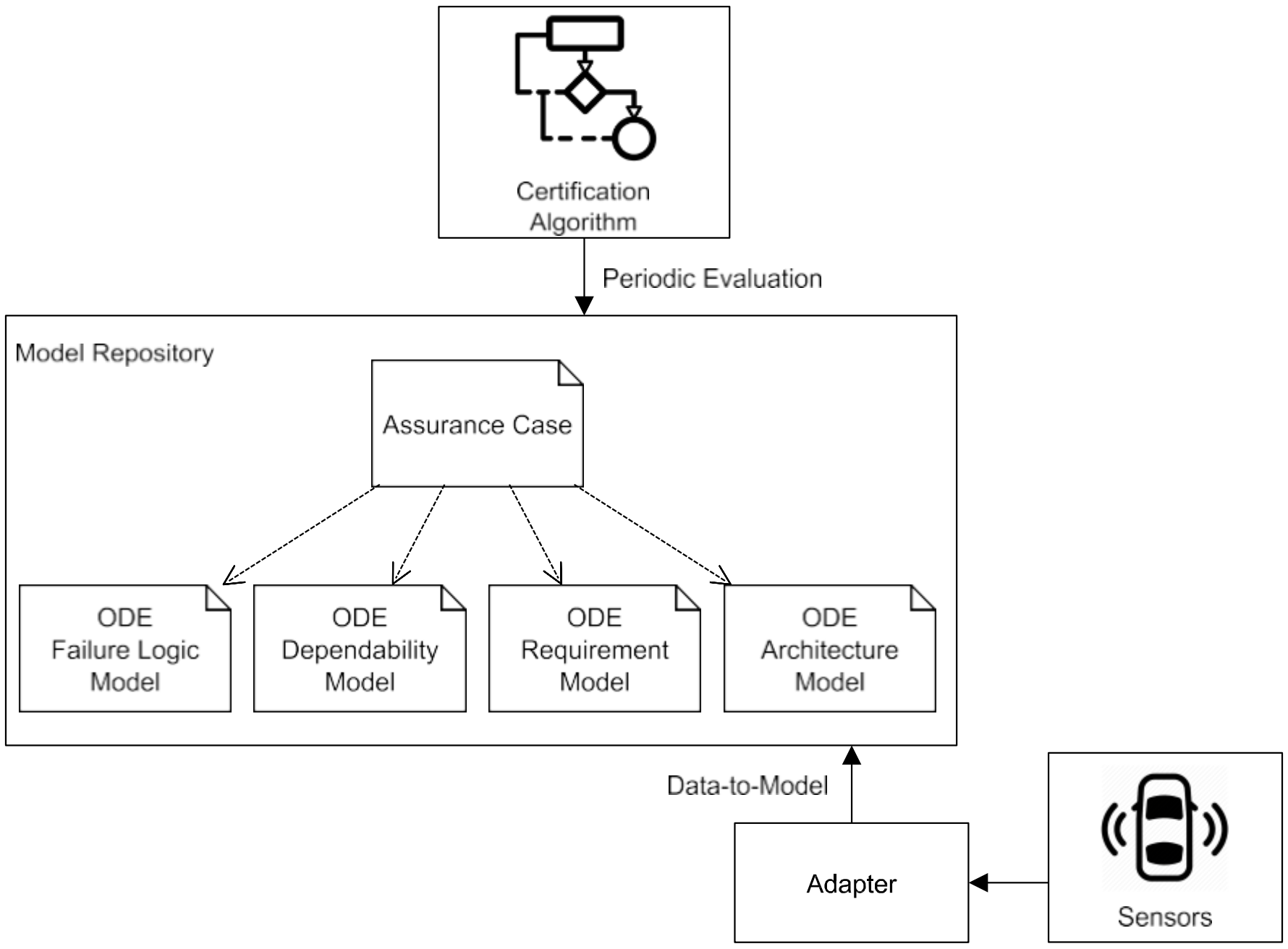

Figure 2: An example of runtime assurance case.

### 2.4. Assurance Cases and the Structured Assurance Case Metamodel

Structured argumentation is also used in other domains, particularly for demonstrating system security [5]. Such augmentations are typically referred to as ***security cases***. The similarities between safety and security cases have been highlighted in [22]. Therefore, the term ***assurance case*** is a broader definition: ***An assurance case should communicate a clear, comprehensible and defensible argument that a system/service is acceptably safe and/or secure to operate in a particular context.***

To promote standardisation and interoperability, the Object Management Group (OMG) specified and issued the ***Structured Assurance Case Metamodel*** (SACM) [28]. SACM is developed by the specifiers of existing system assurance approaches (e.g. GSN and CAE), based on the collective knowledge and experiences of safety and/or security practitioners over the period of last two decades. Therefore, features that are not previously supported by GSN and CAE have been evaluated and included in SACM. A selection of such features are summarised in Table 1 (with an indicative but non-exhaustive list of works that motivate the features).

**Modularity (F1)**. It is important to promote modularity for assurance cases, so that system safety/security can be argued on a component basis (instead of having an enormous assurance case diagram) [9]. Modularity is supported by GSN, with the notion of ***Module*** and ***ContractModule***. In SACM, a finer grained modularity is provided. The users are able to selectively declare the argument/artifact/terminology elements externally through *package interfaces*, and then integrate these packages using *package bindings*. In this way, engineers are able to understand more clearly how assurance cases are integrated.

| Feature | Motivated by |
|---|---|
| F1. Modularity | [9] |
| F2. Multiple Language Support | [8; 7; 25] |
| F3. Controlled Vocabulary | [23; 2] |
| F4. Describing the Level of Trust in Arguments | [13; 10] |
| F5. Counter-Arguments in Assurance Cases | [1] |
| F6. Traceability from Evidence to Artifact | [30; 8] |
| F7. Automated Assurance Case Instantiation | [14; 12] |

Table 1: Features added to SACM

**Multiple Language Support (F2)**. Multiple language support seems trivial in assurance cases created using GSN and CAE. However, when creating SACM version 2.0 we realised that the importance of multiple language support not only lies in the ability to describe arguments in multiple natural languages, but also lies in the ability to describe arguments in computer (e.g. formal) languages [8; 7; 25]. Arguments described using computer languages enable the possibility of (semi-)automated reasoning of system safety, which is particularly important in the context of runtime system assurance.

**Controlled Vocabulary (F3)**. Various studies have identified the importance of controlled vocabulary used in system assurance arguments [23; 2]. In SACM, the users are able to create controlled vocabulary (which can refer to actual model/model elements that define the vocabulary) and refer to them in the assurance argument.

**Describing the Level of Trust in Arguments (F4)**. There is a need to argue the trustworthiness of arguments made in an assurance case, motivated in [13]. In GSN, an extension (*Assurance Claim Points*) was suggested to allow the association of the confidence of arguments to a primary argument. In SACM, there is a facility specifically designed to enable the users to argue the level of trust for argument elements.

**Counter-Arguments in Assurance Cases (F5)**. Sometimes, it is also important to present counter arguments in an assurance case [1]. In GSN and CAE there is no specific means to express counter arguments. In SACM, an assertion can be declared as *counter*, to declare a reversal argument.

**Traceability from Evidence to Artifact (F6)**. In GSN and CAE, evidence in assurance cases are described using natural language, and there is no built-in facility that enables the traceability from evidence to the actual artefact. Existing work achieves traceability through the use of an external metamodel [30; 8]. In SACM, traceability is naturally supported without the need of an external model.

**Automated Assurance Case Instantiation (F7)**. Assurance case templates are useful in capturing good practice in system assurance for re-use. GSN provides the notion of ***GSN patterns***, which enables the users to create abstract safety cases (templates), and then ***instantiate*** the patterns based on actual system information to create concrete safety cases. In [14] and [12], a model-based automated pattern instantiation approach is discussed and presented. This approach uses an intermediate ***weaving model*** to link GSN patterns with system models, and requires extensions to the GSN metamodel. In SACM, automated instantiation can be achieved without the introduction of these extensions.

As principal contributors of SACM and the originators of GSN, we acknowledge that SACM is more powerful than GSN and CAE in terms of expressiveness. Therefore, it is encouraged to use SACM, especially in the context of model-based system assurance. However, in the SACM specification there is limited information on the intended usage of SACM. In order to exploit SACM's full potential, and to promote the adoption of SACM, it is necessary to explain SACM in detail so that safety and security engineers can fully use SACM to achieve higher level goals (e.g. automated model validation to check either well-formedness and/or the validity of runtime assurance certificates).

In the current state of practice, graphical notations such as GSN remain the most popular approach for system assurance. SACM is designed to support GSN, but the OMG specification does not provide a mapping between GSN elements and SACM elements, as there has not been a SACM aligned GSN metamodel. Thus, in this paper we provide a GSN metamodel which aligns to SACM. In addition, there is also a need to translate from GSN to SACM. First of all, the OMG has not defined a concrete syntax (i.e. graphical notation) for SACM elements, which makes creating SACM models a tedious and error-prone process. Thus, to make the transition from GSN to SACM, it is good practice to use GSN notations to construct arguments and then transform to SACM using model-to-model transformation. Secondly, since GSN has been widely adopted in industry, practitioners can convert their legacy diagrams into GSN models, and then transform to SACM to enable model-based system assurance. Due to the reasons above, in this paper we will also provide a mapping (in the form of a model-to-model transformation) from GSN to SACM.

### *2.5. Summarised Motivations*

The motivations of our work can be summarised as follows.

- **The need for model-based assurance case**. As previously discussed, MDE is beneficial to system assurance. High level operations such as model validation, mode-to-model transformation and model merging can be performed on model based assurance cases. In addition, model based assurance cases are the key to assure safety/security related Open Adaptive Systems.

- **Heterogeneous GSN metamodels and misalignment to SACM**. Although a number of model-based assurance case tools exist, they implement their own versions of GSN, and their mappings from GSN to SACM version 1.0 are not unanimous due to the fact that SACM version 1.0 was not sufficiently explained. In addition, due to the release of SACM version 2.0, the claimed supports for SACM of existing model based GSN/CAE tools become out-dated.

- **SACM and its application in Open Adaptive Systems**. SACM provides more features than GSN/CAE, which makes it more powerful in terms of expressiveness. SACM is fundamental to runtime system assurance, which is the key for safety/security related Open Adaptive Systems.
- **Insufficient explanation of SACM**. SACM has been developed over the last decade, and it is the result of significant deliberation and consultation. As a consequence, there are many use cases behind the features (known to the authors of this paper as principal contributors of the SACM standard) that may not be apparent to users on first inspection (e.g. to support existing concepts such as modularity, patterns and meta argumentation). SACM and its relationship with GSN/CAE are not sufficiently explained, which hinders its adoption.

## 3. The Goal Structuring Notation

As SACM is developed based on concepts in the Goal Structuring Notation (GSN), it is necessary to discuss the features and drawbacks of GSN before discussing SACM. GSN is a well established graphical argumentation notation that is widely adopted within safety-critical industries for the presentation of safety arguments within safety cases. The core elements of GSN are shown in Figure 3.

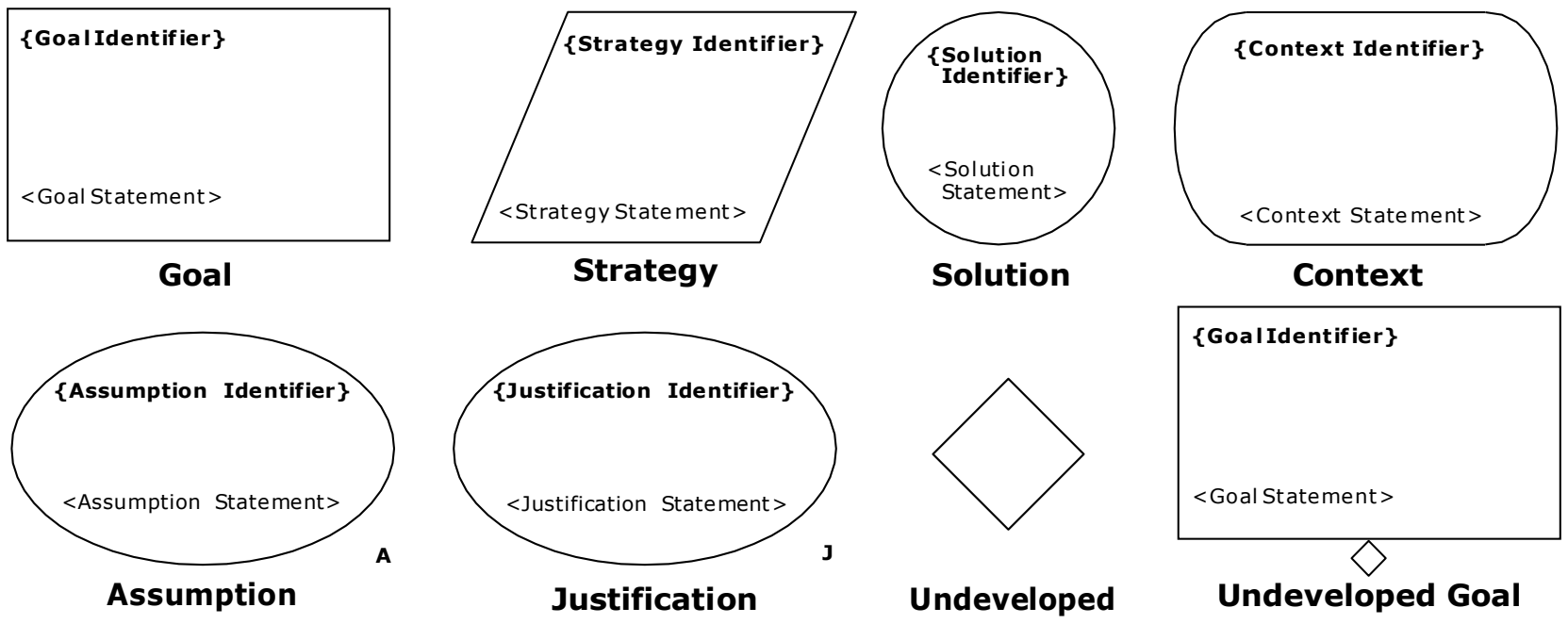

Figure 3: Core GSN elements.

A ***Goal*** represents a safety claim within the argumentation. A ***Strategy*** is used to describe the nature of the inference that exists between a goal and its supporting goal(s). A ***Solution*** represents a reference to an evidence item or multiple evidence items. A ***Context*** represents a contextual artefact, which can be a statement, or a reference to contextual information. An ***Assumption*** represents an assumed statement made within the argumentation. A ***Justification*** represents a statement of rationale.

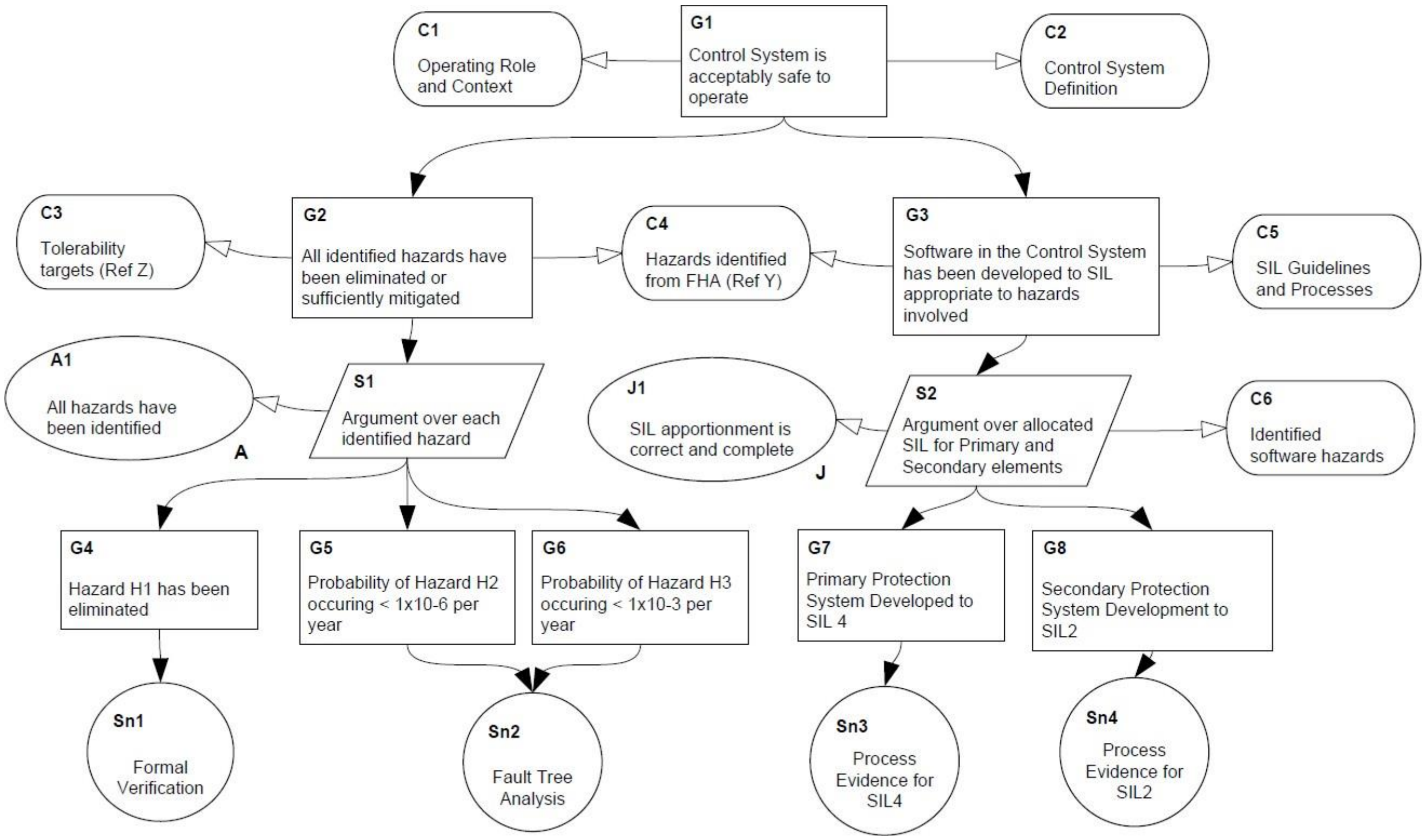

Figure 4: An example of a goal structure.

An element can be ***Undeveloped***, which means that a line of argument has not been developed yet (meaning it being abstract and needs to be instantiated). The ***Undeveloped*** notation can apply to ***Goal***s and ***Strategies***. The ***Undeveloped Goal*** in Figure 3 is an example.

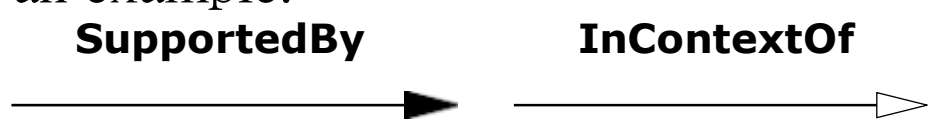

Figure 5: GSN connectors.

Core elements of GSN are connected with two types of connectors, as shown in Figure 5. The ***SupportedBy*** connector allows inferential or evidential relationships to be documented. The ***InContextOf*** relates contextual elements (i.e. ***Context***, ***Assumption*** and ***Justification***) to ***Goal***s and ***Strategies***.

When elements of GSN are linked together in a network, they are often referred to as a ***goal structure***. The purpose of a goal structure is to show how ***Goal***s are successively broken down into sub-***Goal***s until a point is reached where ***Goal*** s can be supported by direct reference to available evidence (***Solution***s). An example of a goal structure is shown in Figure 4[5].

Goal structures can be organised in ***Module***s. For example, for a system that consists of two components A and B, it is possible to organise the safety cases of component A and B in two Modules ***MA*** and ***MB***. Modularity promotes re-use, so that safety cases for system components can be re-used when different components are integrated to form a system. Figure 6 shows the GSN elements that enable modularity support. When integrating system safety cases, a ***Contract Module*** can be used to ***bind*** different ***Module***s together.

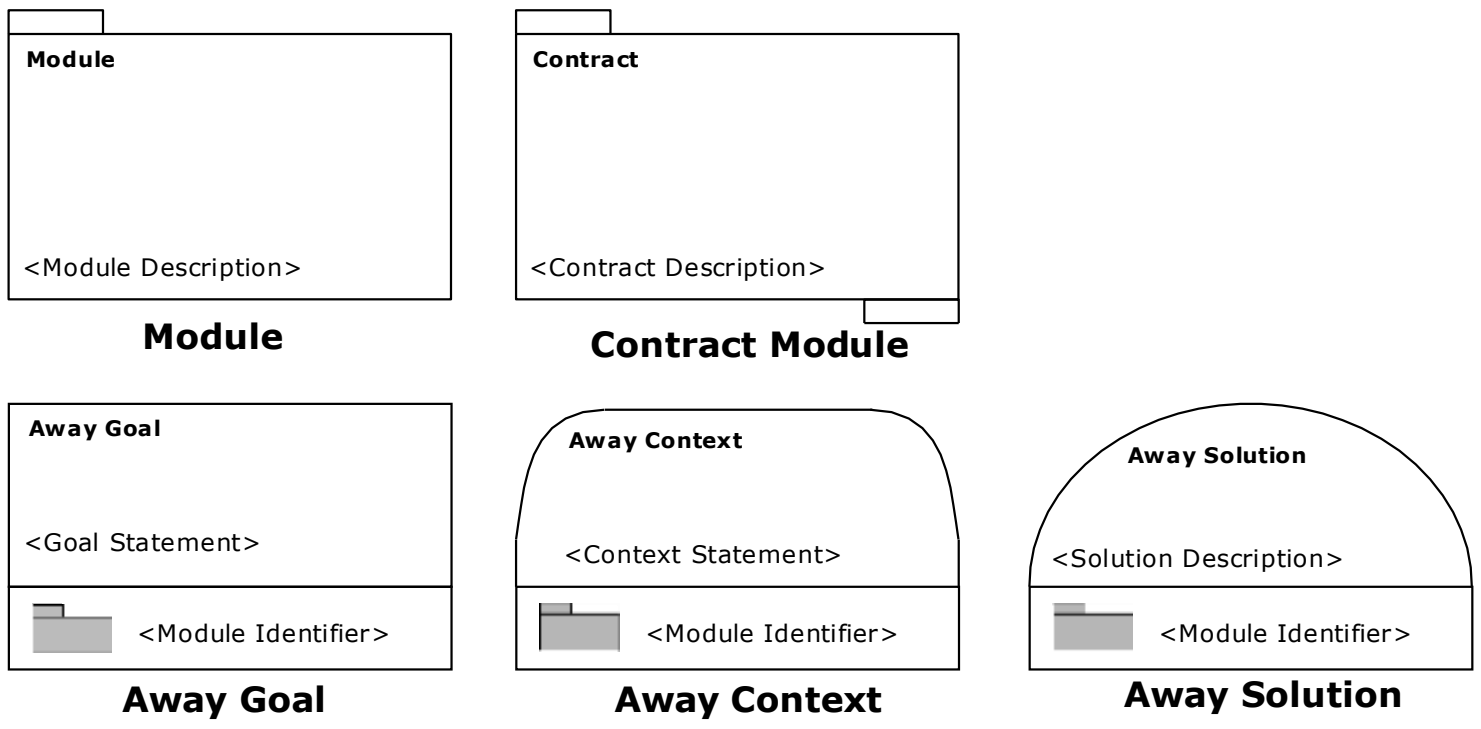

Figure 6: Modular GSN elements.

Binding is done via the use of ***Away Goal*** s, ***Away Context*** s and ***Away Solution***, where ***Goal*** s, ***Context*** s and ***Solution***s from an external ***Module*** can be referenced. Like other GSN elements, ***away*** elements can be connected using ***SupportedBy*** and ***InContextOf*** connectors.

[5]From the GSN standard: https://www.goalstructuringnotation.info/

For a successful GSN safety case, safety engineers tend to define a template of the safety case to re-use its structure in the future. GSN provides the extension for users to define templates that are called ***GSN patterns***. In [19], the use of patterns is discussed, pattern are a means of documenting and reusing successful assurance argument structures. Safety case argument patterns provide a way of capturing the required form of a safety argument in a manner that is abstracted from the details of a particular argument. It is then possible to use the patterns to create specific arguments by ***instantiating*** the patterns in a manner appropriate to the application. Pattern instantiation refers to the process of constructing concrete system safety cases by populating the templates provided in the GSN pattern with actual system information. Figure 7 shows the elements in GSN which enable the users to create patterns.

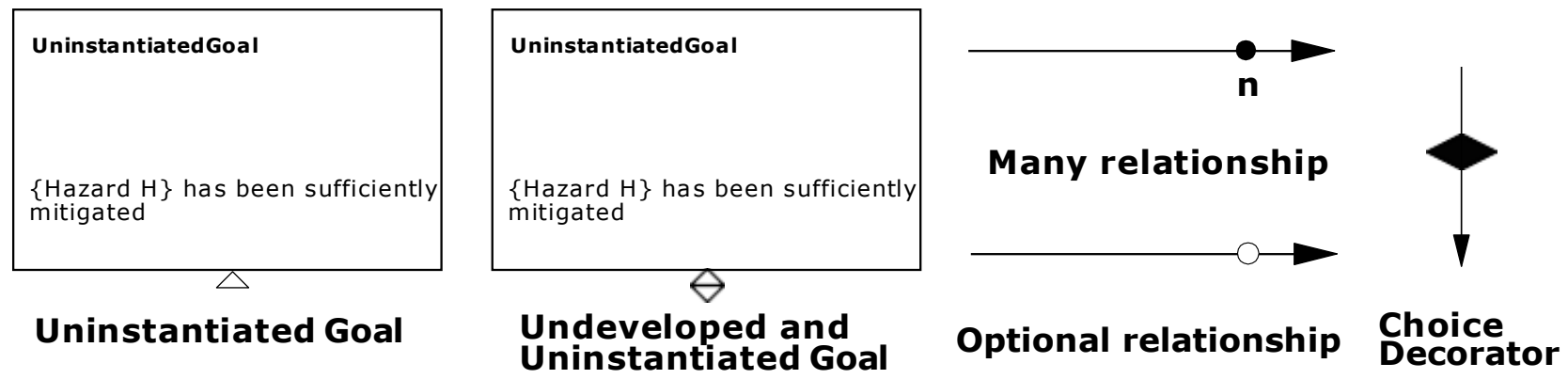

Figure 7: GSN pattern elements.

The ***Uninstantiated*** indicator marks that an element is abstract and needs to be instantiated. At some later stage, the abstract element needs to be replaced with a more concrete instance. ***Uninstantiated indicator*** can be associated to any GSN element. Figure 7 demonstrates how it can be associated to a ***Goal***. The ***Undeveloped and Uninstantiated*** indicator marks an element (in particular, ***Goal*** s and ***Strategies***) to be both abstract (to be instantiated) and needs more development (needs supporting argument), Figure 7 demonstrates its usage on a Goal. In GSN patterns, the ***SupportedBy*** and ***InContextOf*** connector can bear more information, the ***Many*** decorator on a connector indicates that when the pattern is instantiated, the connector can be multiplied ***n*** times (expressed in the label) based on the actual system information provided. The ***Optional*** decorator indicates that when the pattern is instantiated, the connector can connect to one or zero element. The ***Choice*** decorator on a connector can be used to denote possible alternatives in satisfying a relationship, which can represent a 1-of-n or m-of-n selection.

Figure 8 shows an example of a GSN pattern (adopted from [19]). The contents in the curly brackets are ***role***s in GSN terms, they are place holders which will be replaced by actual system information when the pattern is instantiated. For example, {System X} in G1 will be replaced with the actual name of the system when the pattern is instantiated. Similarly, the ***SupportedBy*** connector between S1 and G2 specifies that when the pattern is instantiated, there should be $n$ (the number of safety-related functions implemented by the system) ***Goal***s and $n$ ***SupportedBy*** connected to S1. Pattern instantiation is often a manual process that involves comprehension of the GSN patterns and replacing the ***role***s with actual system information. There has been work on automating the pattern instantiation process using MDE with the use of a ***weaving model*** [12]. In this section, we briefly discussed the elements provided by GSN. GSN is powerful in representing arguments in a structured way, which enables better comprehension of system safety arguments. GSN promotes modularity and abstraction in the sense that good practice in safety case construction can be captured and re-used. In the following sections, we will discuss SACM and its relationship with GSN.

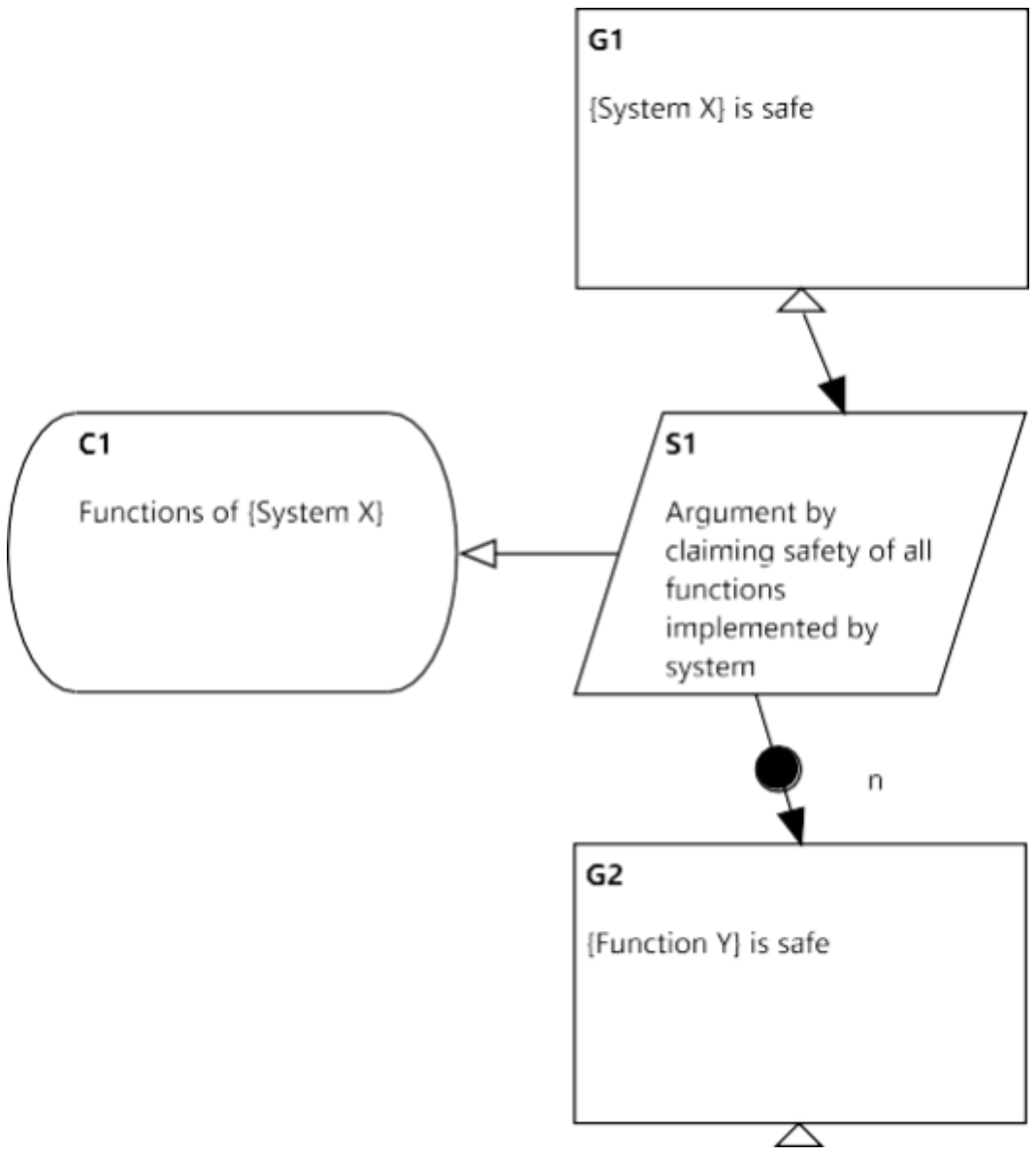

Figure 8: An example of a GSN Pattern [19]

## 4. Structured Assurance Case Metamodel

The ***Structured Assurance Case Metamodel*** (SACM) is standardised by the Object Management Group (OMG). The intention of the metamodel is to promote a model-based approach in the process of system assurance, which is currently a manual approach that produces artefacts (i.e. Assurance Cases) that are (mostly) not model-based, where higher level operations (such as model validation) on these artefacts are not applicable. SACM is created to support model-based paradigm with existing well established graphical argumentation notations, the ***Goal Structuring Notation*** (GSN) and ***Claims-Arguments-Evidence*** (CAE).

In this section, we discuss the packages in SACM and explain their intended usage. Two types of examples are provided to demonstrate the usage of SACM: for simple concepts to explain without further context, we use in-place examples; and for complex concepts which requires the context of understanding the entirety of SACM, we use concrete examples provided at the end of this section. Since OMG has not standardised the concrete syntax of SACM[6], we use object diagrams in the examples.

### 4.1. SACM Overview

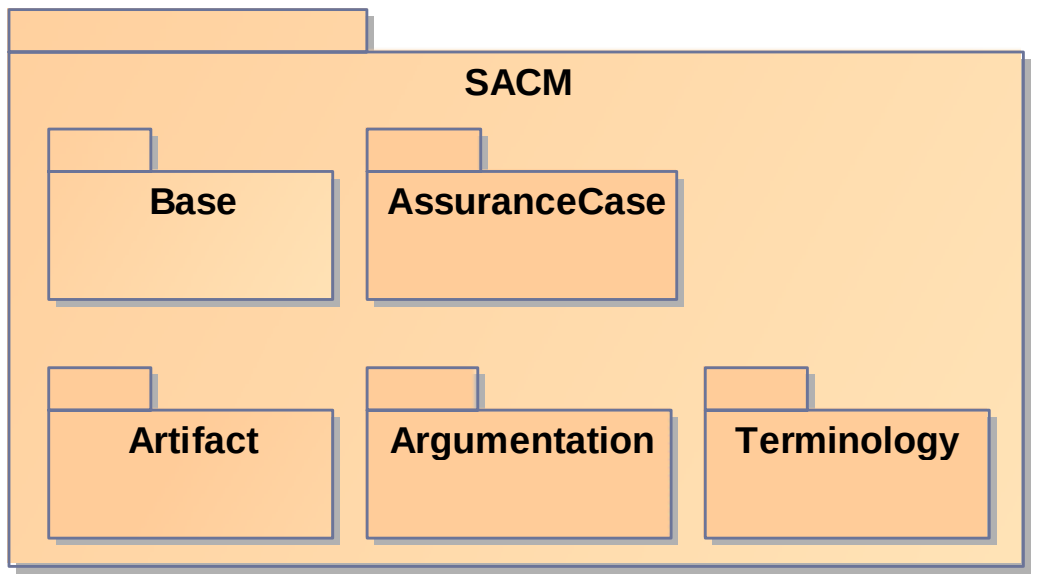

Figure 9: Components of SACM

There are five components in SACM, as shown in Figure 9. The ***AssuranceCase*** component (discussed in Section 4.2) captures the concepts of ***Assurance Case*** in system assurance. In SACM, an ***AssuranceCase*** package contains a number of ***Argumentation*** packages, ***Terminology*** packages and ***Artifact*** packages. The ***Base*** component (discussed in Section 4.3) defines the fundamental elements of SACM, such as element names and descriptions. The ***Artifact*** component (discussed in Section 4.4) captures the concepts used in providing evidence for the arguments made for system properties. The ***Terminology*** component (discussed in Section 4.5) captures the concepts used in expressing the arguments regarding system properties, such as ***Expression***s. The ***Argumentation*** component (discussed in Section 4.6) captures the concepts used in arguing system properties (such as safety and/or security)[7].

### 4.2. SACM AssuranceCase Component

The ***AssuranceCase*** component provides an insight on how an ***Assurance Case*** in SACM is organised, hence it is discussed first. The structure of the ***AssuranceCase*** component is shown in Figure 10. The core element in the ***AssuranceCase*** component is the ***AssuranceCasePackage*** element, which extends the ***ArtifactElement*** in the ***Base*** component. The implication is that an ***AssuranceCasePackage*** can be considered to be an artifact. An ***AssuranceCasePackage*** can hold a number of ***ArgumentPackage***s, ***TerminologyPackage***s and ***ArtifactPackage***s, which holds the argumentation with regards to system safety/security, the controlled vocabularies used in the argumentation, and the artifact that backed the argumentation as evidence, respectively. In this way, SACM provides more detailed support for modularity than GSN[8].

[6]We are aware that concrete syntax for SACM is in development.

[7]System properties hereby refer to the safety and/or security in the context of this paper.

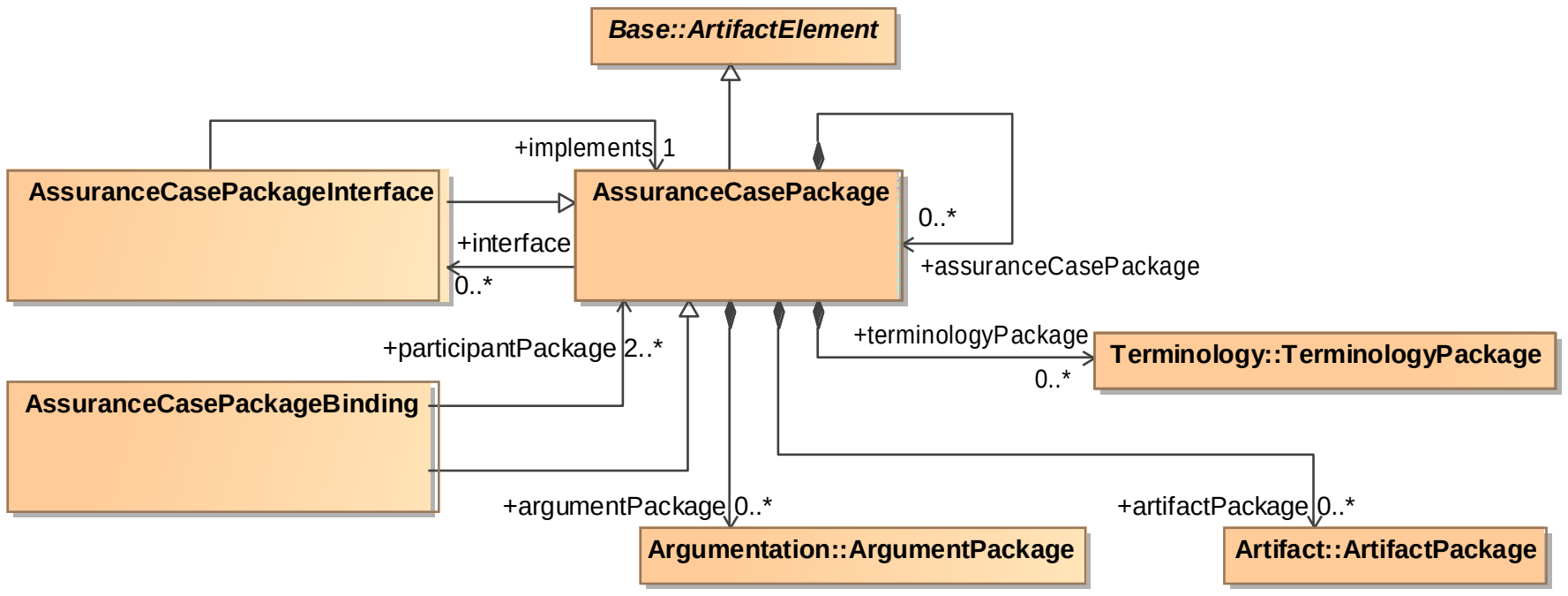

Figure 10: SACM AssuranceCase Component

Sometimes the developer of an ***AssuranceCasePackage*** may want to make part of the ***AssuranceCasePackage*** available externally so that they can be reused. Consider the scenario where a system is composed of components A and B. ***AsssuranceCasePackage***s ACPa and ACPb are created respectively for A and B (which contain structured argumentations with regard to system properties for A and B). The developer may want to make parts of the argumentations public so that during system integration, where A and B are integrated to form a system, their assurance cases ACPa and ACPb can also be integrated to form a new ***AssuranceCasePackage***. To disclose only necessary contents externally, one needs to make use of the ***AssuranceCasePackageInterface***. The premise of system integration in safety related domains is to integrate assurance cases of systems to form an overall assurance case. SACM handles this scenario with the ***AssuranceCasePackageBinding***, which binds two or more ***AssuranceCasePackageInterface***s together to form an overall ***AssuranceCasePackage***. This particular scenario is discussed in Section 5.4.

### *4.3. SACM Base Component*

The ***Base*** component captures the fundamental concepts of SACM, the structure of the ***Base*** package is shown in Figure 11. The base element of all SACM elements is ***Element***. Its direct children are ***LangString***, ***MultiLangString*** and ***SACMElement***. ***LangString*** is used as an equivalence to ***String*** (the value of which is held in the ***+content*** feature), except it captures an additional feature

[8]Feature **F1** in Section 2.4.

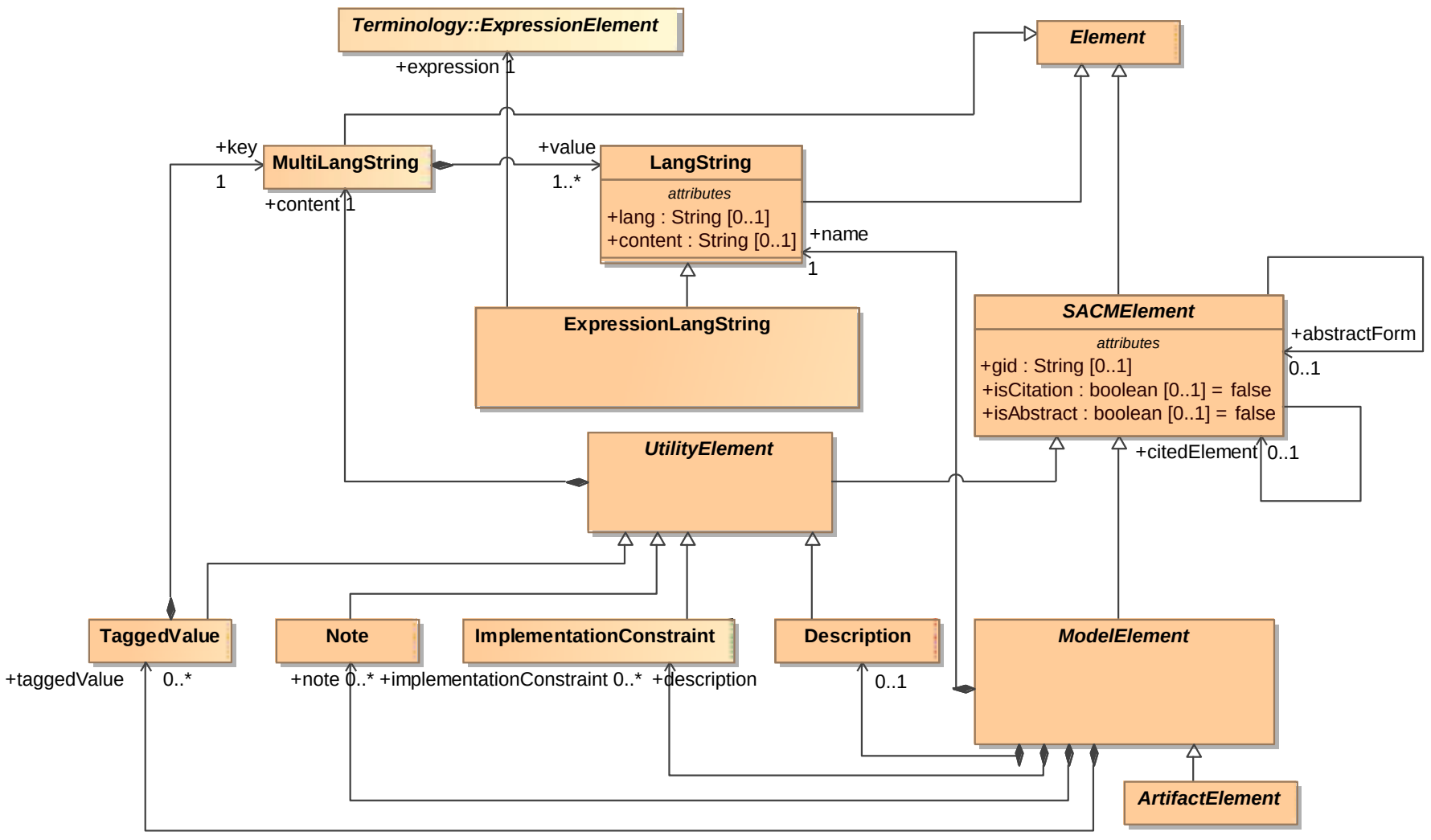

Figure 11: SACM Base Component

*+lang* which allows the users to define what language is used in the *LangString*. *ExpressionLangString* is used to not only record a *String* in SACM, but also refer to its corresponding *Expression* organised in a *TerminologyPackage*. The usage of *ExpressionLangString* is discussed in Section 4.5. *MultiLangString*, as its name suggests, is used to express the same semantics using different languages. For example, to express 'hazard' in both English and German, the user can create a *MultiLangString* with two *LangString*, as shown in Figure 12.

*MultiLangString* can be associated to SACM elements to denote the same meaning. What is more important than multiple natural language support is the support for computer languages[9]. As previously discussed, for open adaptive systems, system assurance needs to be performed at runtime. Hence, automation is needed at runtime to reason about the safety of open adaptive systems on assurance models. A first step towards this direction is the use of formal languages in assurance cases, so that system assurance can be (semi-) automated, in the sense that automated reasoning can be performed on the argumentation. In this case, *MultiLangString* can be used to hold both natural languages and computer languages (e.g. formal languages) to support automated reasoning of argumentations.

*SACMElement* contains the foundational features of all SACM elements. *SACMElement* can record a *+gid* (*global identification*). *SACMElement* is also able to refer to (or 'cite') other *SACMElement* s, which is useful for implicit references, discussed in Section 5.1. The *+citedElement* and *+isCitation* prop-erties are used for this purpose. A ***SACMElement*** can also be abstract, denoted by the *+isAbstract* property, and can be an *+abstractForm* of another, which is discussed in Section 5.3.

[9]Feature **F2** in Section 2.4.

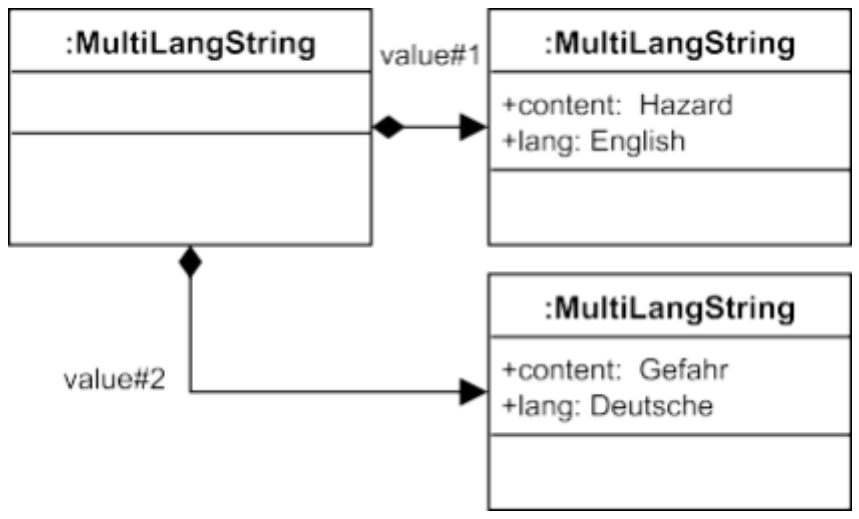

Figure 12: Example *MultiLangString*.

***ModelElement*** further refines ***SACMElement*** which contains a ***name*** (of type ***LangString***) and a set of ***UtilityElement*** s. A ***ModelElement*** can contain a ***Description*** which describes its contents. As previously mentioned, a ***Description*** can be expressed in any language via the usage of ***MultiLangString***. A ***ModelElement*** can also contain an ***ImplementationConstraint***, which is used to specify the instantiation rules for assurance case templates (that of similar to safety case patterns). A ***ModelElement*** can also contain a number of ***Note***s, to hold additional information other than descriptions. Finally, a ***ModelElement*** can contain a number of ***TaggedValue***s, which are essentially {key, value} pairs. ***TaggedValue*** can be considered as an extension mechanism to allow the users to associate additional features to a ***ModelElement*** (other than the features modelled in the current version of SACM).

The ***Base*** component also defines the ***ArtifactElement***, in the sense that all elements that extend ***ArtifactElement*** are considered to be ***Artifact***s. The reason for this is discussed in Section 5.2.

In summary, the ***Base*** component defines the foundation of SACM. It provides facilities to express assurance cases (to be precise, assurance case models) in natural language, as well as in computer languages. The ***Base*** component also provides a number of ***UtilityElement*** s so that the user can use them to describe ***ModelElement*** s as precisely as possible.

### *4.4. SACM Artifact Component*

Before delving into the ***Argumentation*** component, it is necessary to discuss the ***Artifact*** component. The structure of the ***Artifact*** component is shown in Figure 13. All elements in the ***Artifact*** component extend ***ArtifactElement*** in the ***Base*** component. ***ArtifactElement***s are organised in ***ArtifactPackage***s to promote modularity. Assurance case integration needs also be performed at the ***ArtifactPackage*** level, which is fulfilled by ***ArtifactPackageInterface*** and ***ArtifactPackageBinding***. ***ArtifactGroup*** is a new concept introduced in SACM 2.0. As ***ArtifactPackage*** can contain rather a large number of ***ArtifactElement***s, the

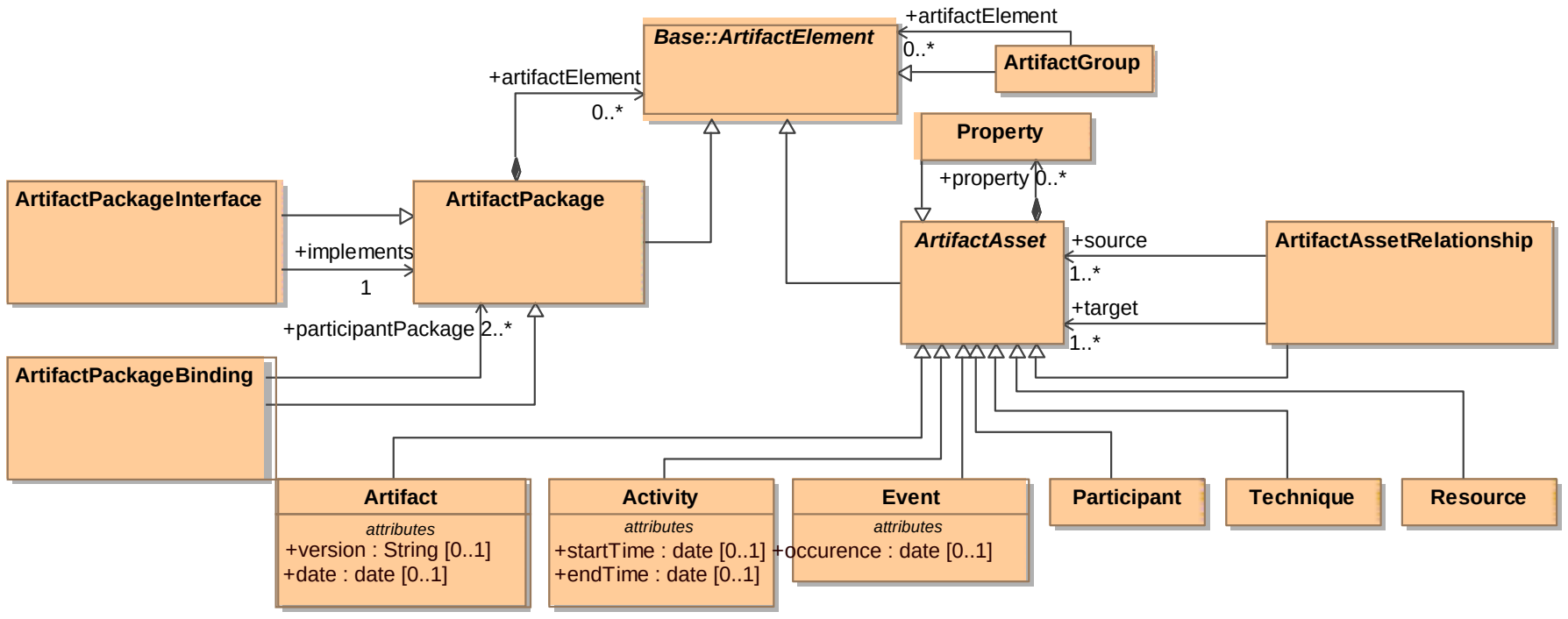

Figure 13: SACM Artifact Component

***ArtifactGroup*** provides the user with a means to selectively group ***ArtifactElements***, so that the user can group/view ***ArtifactElement*** s with their defined criteria.

***ArtifactAsset*** allows the users to create corresponding artefact elements in SACM, it can contain a number of ***Property***-ies to hold user-defined properties. ***Artifact*** records a piece of information (e.g. hazard log, failure logic models, etc). ***Activity*** records an activity (e.g. specification of requirements). ***Event*** records an event (e.g. creation/modification of ***Artifact*** s). ***Participant*** records participants involved in ***ArtifactAsset***s (e.g. safety engineers). ***Technique*** records the techniques used in ***Activities*** (e.g. requirement elicitation). ***Resource*** records a piece of resource, usually in the form of some electronic files. And finally, ***ArtifactAssetRelationship*** is used to link ***ArtifactAsset***s (e.g. connecting ***Activity*** to ***Participant*** s). Note that the ***ArtifactAssetRelationship*** is a generic relationship, however, the user can choose to use ***Property*** to specify the purpose of an ***ArtifactAssetRelationship***.

One open question regarding the ***Artifact*** component is how to refer to external materials (such as system requirements model, system design model, system failure logic model, etc.). To achieve automation, it is necessary to have a means to refer to external materials (especially models) so that the assurance case can be validated automatically with its supporting evidence. The user may make use of the ***Property*** element, as shown in Figure 14, where a ***Property*** is associated to an ***Artifact***[10], which has a ***name*** (of type ***LangString***): 'URI' and a ***Description***, which in turn specifies a file on local disk (systemdesign.model). In this way, the ***Property*** element is used as a reference to a local file.

SACM does not restrict how external materials should be referenced, the description provided above is one way of achieving it. It also depends on tool implementations on how external references is handled. It is possible to have finer grained reference to (a collection) of model elements for a model (e.g. a Fault Tree Analysis model). However, OMG has not standardised how this can be done.

[10]The ***Artifact*** should have its own features such as name and description, which is omitted here.

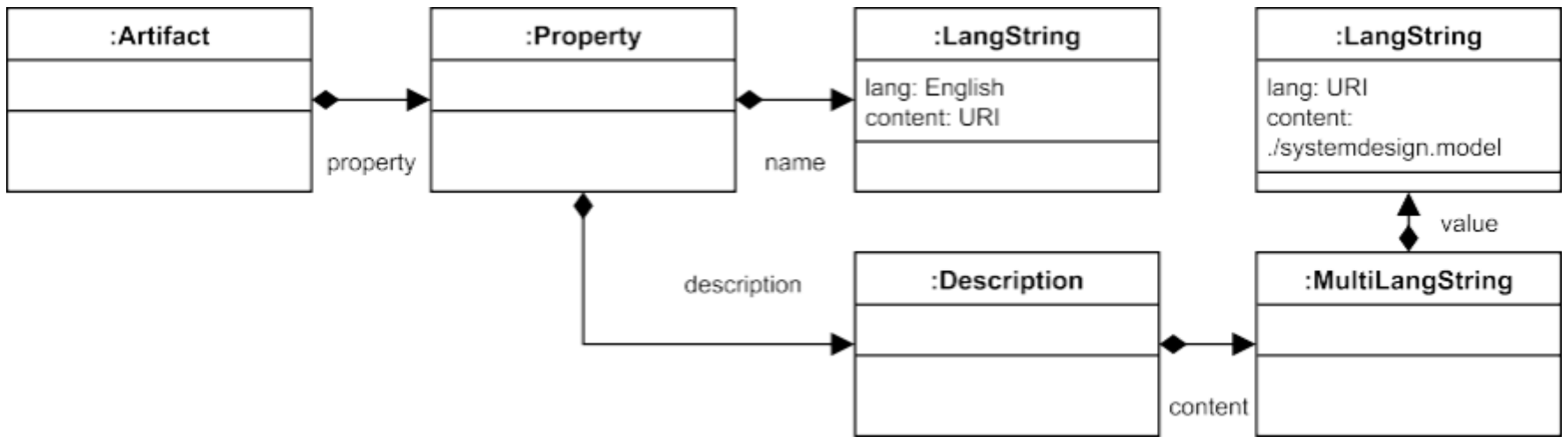

Figure 14: External reference using *Property*.

## *4.5. SACM Terminology Component*

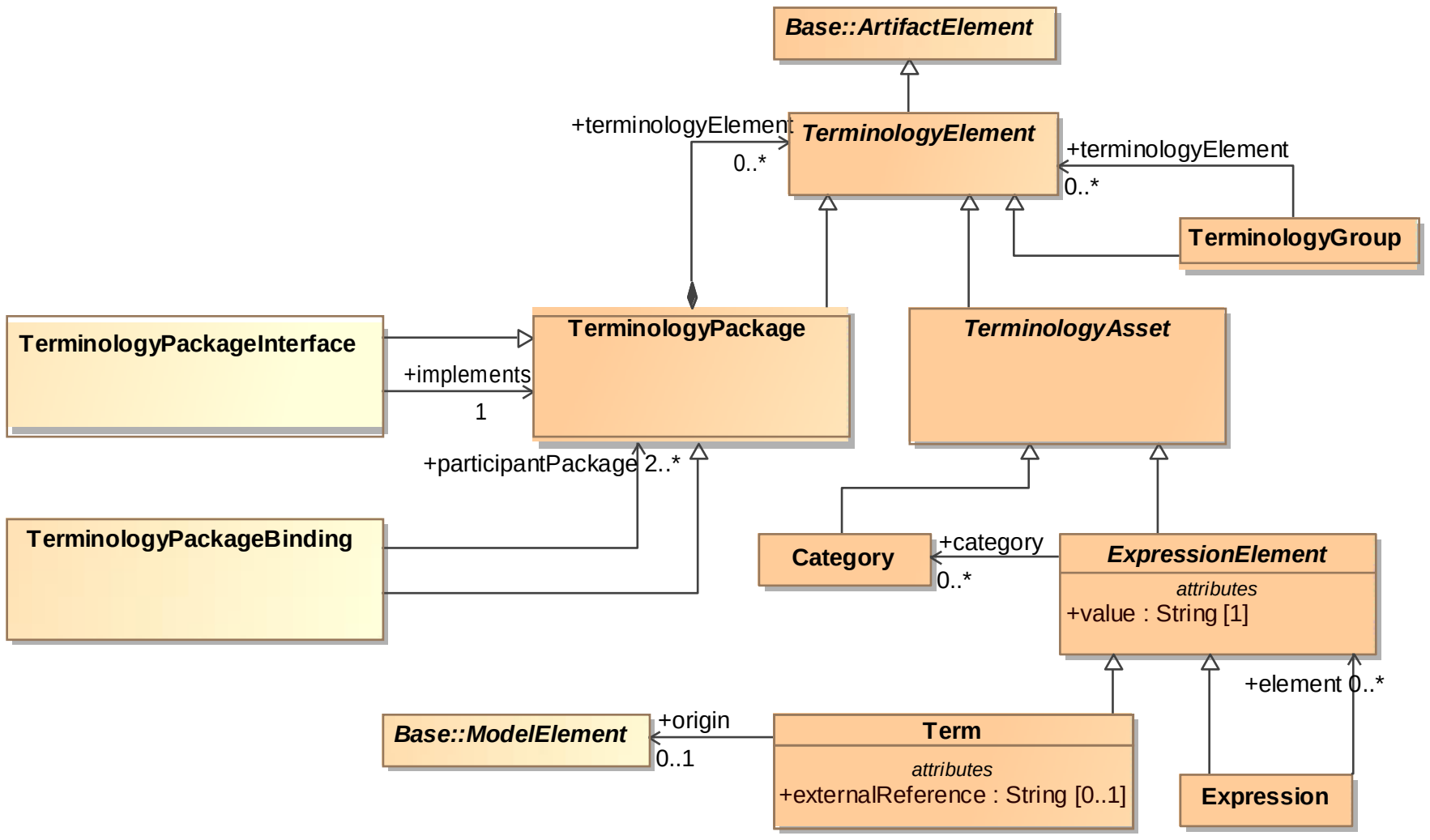

Figure 15: SACM Terminology Component

The *Terminology* component captures concepts that enable the users to define controlled vocabularies[11] to describe their argumentation with greater precision. The structure of the *Terminology* package is shown in Figure 15. The root element of the *Terminology* component is *TerminologyElement* which also extends *ArtifactElement* in the *Base* package. *TerminologyAsset* s are organised in *TerminologyPackage*s. *TerminologyInterface* and *TerminologyPackageBinding* are used for assurance case integration. *TerminologyGroup* enables to users to group *TerminologyAsset* s selectively.

[11]Feature **F3** in Section 2.4.

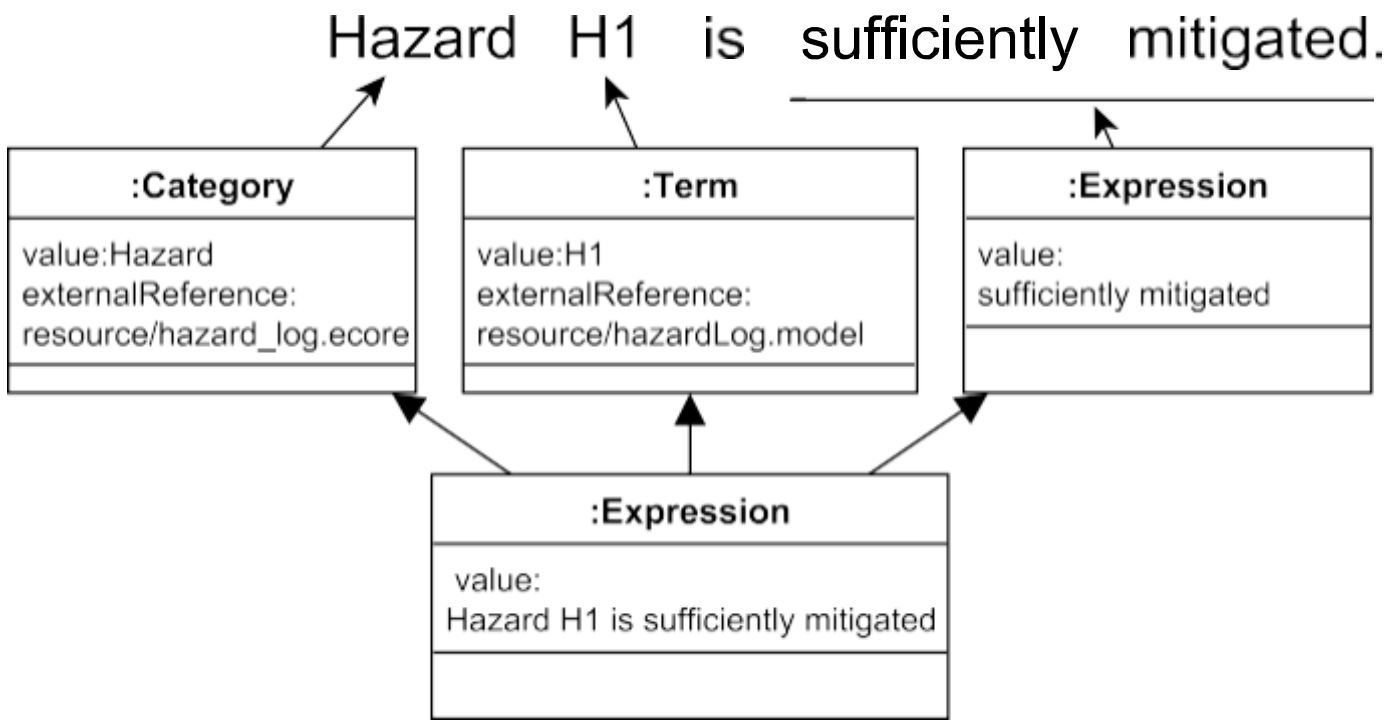

Figure 16: Example use of elements in the Terminology component.

*ExpressionElement* captures two expression concepts used in argumentations: *Expression* and *Term*. An *Expression* is used to model phrases and sentences, and a *Term* is used to define a terminology used in the argumentation. In addition to *ExpressionElement*, there is also *Category* which is used to group *Term*s and *Expression*s into categories. An example is provided in Figure 16. In the sentence **Hazard H1 is sufficiently mitigated**, *Hazard* is defined as a *Category* (for it defines the concept of ' Hazard' in the hazard log.ecore metamodel through external reference), the word *H1* is defined as a *Term* as it refers to a specific hazard in the hazard log model (hazardLog.model), the phrase *sufficiently mitigated* is an *Expression* as it is commonly used phrase in system assurance.

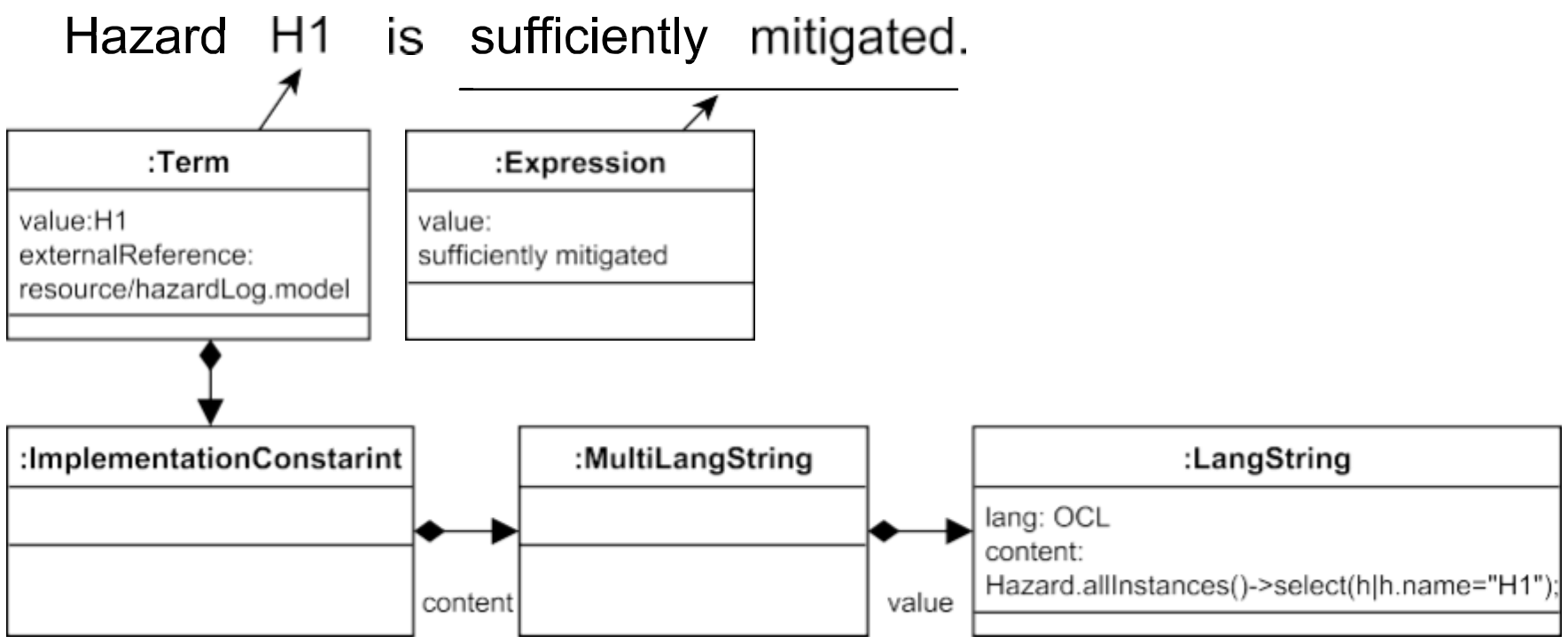

Figure 17: Example use of *Expression* and *Term*.

Whilst ***Term*** element is used to capture a terminology used in the process of system assurance, it can also relate to model elements in other models. For example, in Figure 17, in the statement: **Hazard <u>H1</u> is sufficiently mitigated.**, <u>H1</u> may cross-reference to an identified hazard in a hazard log. The location of the hazard log is recorded in the ***externalReference*** of the ***Term***. In addition, the ***Term*** has an ***ImplementationConstraint***, which allows the users to define model queries to obtain model element(s). In this example, the Object Constraint Language (OCL) is used to obtain the single model element ***H1*** in the hazard log. Again, SACM does not restrict how external references should be handled, and the description provided above is one way of achieving it. Sometimes, a term may refer to a ***ModelElement*** within the assurance case (via the ***+origin*** feature). This will be discussed in Section 4.6.

### 4.6. SACM Argumentation Component

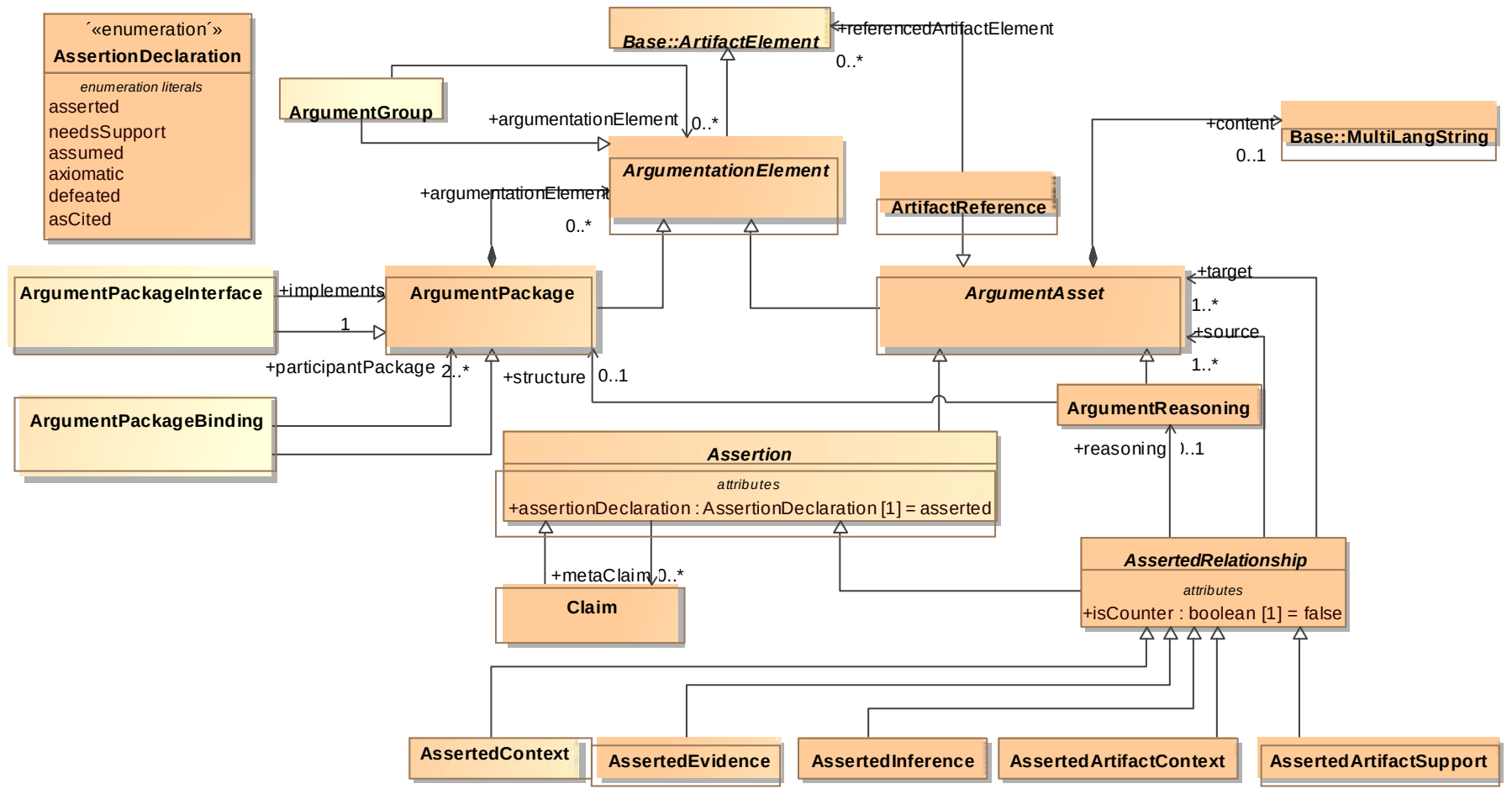

Figure 18: SACM Argumentation Component

The ***Argumentation*** component captures the concepts required to model structured arguments regarding system properties. The structure of the ***Argumentation*** component is shown in Figure 18. The root element of the ***Argumentation*** component is the ***ArgumentationElement***, which is a direct child of ***ArtifactElement*** in the ***Base*** component. This implies that all elements in the ***Argumentation*** package are also considered to be artefacts.

***ArgumentationPackage*** can contain a number of ***ArgumentationElements***, which can either be ***ArgumentPackage***s (and their children) or ***ArgumentAsset*** s. ***ArgumentAsset*** can store a ***content***, discussed in Section 4.3. The content can be in any language[12].

[12]Via the usage of ***MultiLangString***.

*ArgumentAsset* and its children are the elements that form the structured argumentation in the *Argumentation* package. An *Assertion* has an *AssertionDeclaration* to distinguish different types of *Assertion*s. The *AssertionDeclaration*s are as follows:

- **asserted** - the default declaration, means that the *Assertion* is made and is supported by evidence;
- **needsSupport** - a flag indicating that the *Assertion* is not supported yet (needs further development);
- **assumed** - a flag indicating that the truth of the *Assertion* is assumed although no supporting evidence is provided;
- **axiomatic** - a flag indicating that the truth of the *Assertion* is axiomatically true without further supporting evidence;
- **defeated** - a flag indicating that the truth of the *Assertion* is invalidated by a counter-evidence and/or an argumentation;
- **asCited** - when an *Assertion* 'cites' another *Assertion* (via the +citedElement property in *SACMElement*), the truth of the *Assertion* is derived from the value of the cited *Assertion*.

The use of *AssertionDeclaration* is illustrated in Section 5.1.

Sometimes it is necessary to argue the confidence (level of trust) in the arguments provided in the assurance case[13]. This can be achieved using the *+metaClaim* feature of the *Assertion* element. *+metaClaim* as its name suggests, is a *Claim* about an *Assertion* to argue the soundness/trustworthiness of the *Assertion*. Within the meta *Claim*, one may write 'full confidence in Claim C1 is achieved via...' in its description, within which the *Term* C1 refers to another *Claim* in the same SACM model. This is where the *+origin* feature of the *Term* is used, in the sense that the *Term* C1 refers to a *Claim* C1 within the same SACM model.

*Claim* and *AssertedRelationship* are the elements that form structured argumentation. *AssertedRelationship* is used to connect *ArgumentAsset*s to form structured argumentation (*AssertedRelationship*s can also be *counter* relationships, indicated by the *+isCounter* feature to present counter-arguments[14]):

- **AssertedContext** - this relationship is used to connect contextual *Assertion*s to an *Assertion*;
- **AssertedEvidence** - this relationship is used to connect evidence (referenced via *ArtifactReference*) to an *Assertion*;

[13] Feature **F4** in Section 2.4.

[14] Feature **F5** in Section 2.4.

- **AssertedInference** - this relationship records the inference between one or more *Assertion*s and another *Assertion*;
- **AssertedArtifactContext** - this relationship is used to connect contextual artifacts (via *ArtifactReference*) to an artifact (via *ArtifactReference*);
- **AssertedArtifactSupport** - this relationship is used to connect supporting artefacts (via *ArtifactReference*) to an artifact (via *ArtifactReference*).

*ArtifactReference* is a type of *ArgumentAsset*, which is able to refer to an *ArtifactElement*. *ArtifactReference* is typically used to refer to evidence stored in an *Artifact* package. In addition, it can refer to any element that extends *ArtifactElement* (all elements in the *AssuranceCase*, *Argumentation*, *Artifact* and *Terminology* packages are *ArtifactElement*s)[15].

*ArgumentReasoning* is also a type of *ArgumentAsset*. It is used to provide an explanatory description for an *AssertedRelationship*. A detailed discussion of *ArgumentReasoning* is in Section 5.2.

### 4.7. Summary

In this section, we discussed the components provided by SACM. We explained the intended semantics of the elements in the packages and we used some in-place examples to illustrate how SACM can be used to construct a system assurance case with argumentations regarding system properties (i.e. safety and/or security) and its supporting evidence/artifact (using the *Artifact* and *Terminology* packages of SACM). In the next section, we will illustrate how SACM can be used with more concrete examples.

## 5. SACM: Examples

In this section, we provide concrete examples on using SACM to construct structured argumentation, and to form argumentation patterns (similar to GSN argument patterns). Since GSN notations are widely accepted and understood, we compare GSN depictions with their equivalent model elements in SACM to illustrate how SACM elements can be used to denote the same semantics carried by their GSN counterparts.

### 5.1. Example: Making Claims and Citations

Figure 19 provides an example on how to construct a *Claim*. A *Claim* can have a name and a description, captured by *LangString* and *Description* respectively. A *Claim* is **asserted** if it is supported by other *Claim*s. In this example, a claim is connected by an *AssertedInference* with another *Claim* (details ommitted). The GSN equivalent of *Claim* C1 in Figure 19 is provided in

[15] Feature **F6** in Section 2.4.

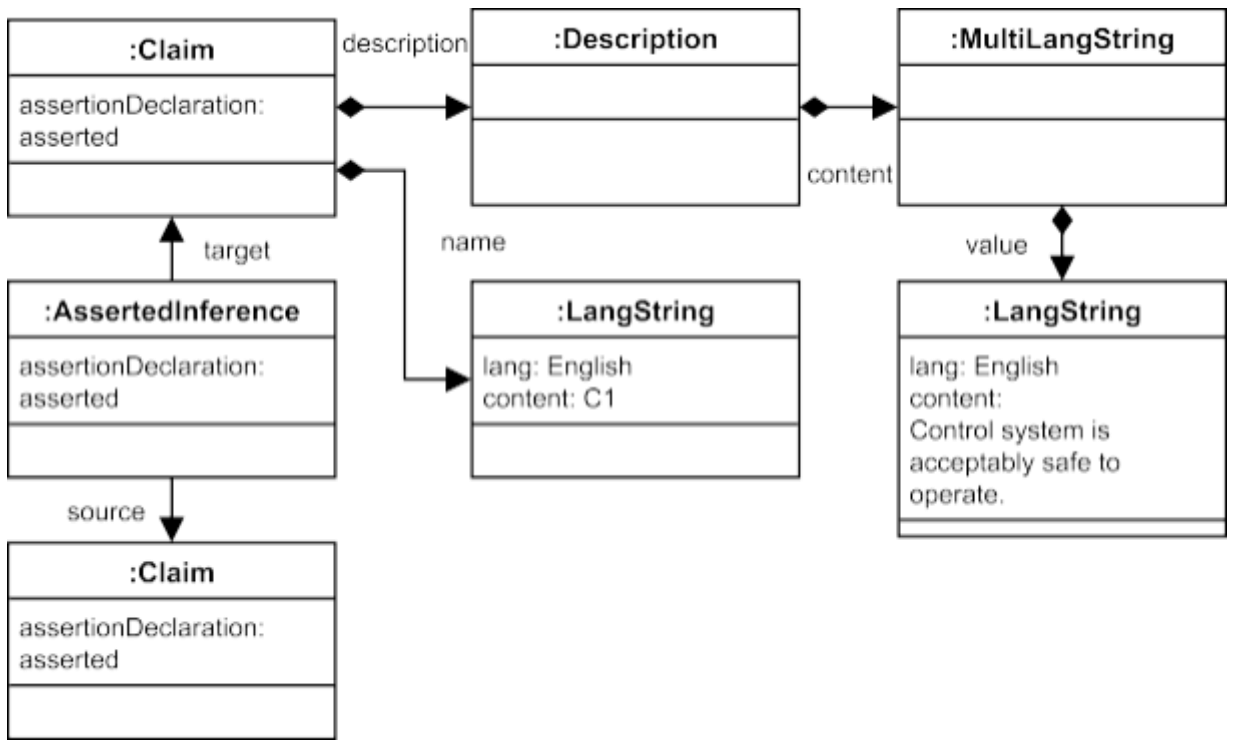

Figure 19: An **asserted** *Claim* in SACM.

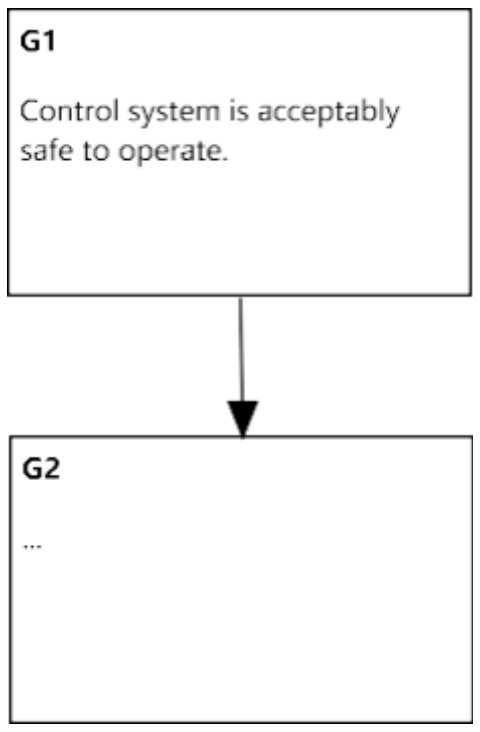

Figure 20: A *Goal* structure in GSN.

Figure 20, where the *Goal* G1 bears the same semantics with C1 (and how *Goal* G1 is supported by another *Goal* [16]).

As previously mentioned, a *Claim* can be marked as **needsSupport**. Figure 21 illustrates the use of this declaration. A *Claim* that is not supported by any argument/evidence should be marked as **needsSupport**. The equivalence of this *Claim* in GSN is shown in Figure 22, where *Goal* G1 bears the same semantics of *Claim* C1. When a *Goal* is not supported by argument/evidence, it is marked as *undeveloped*.

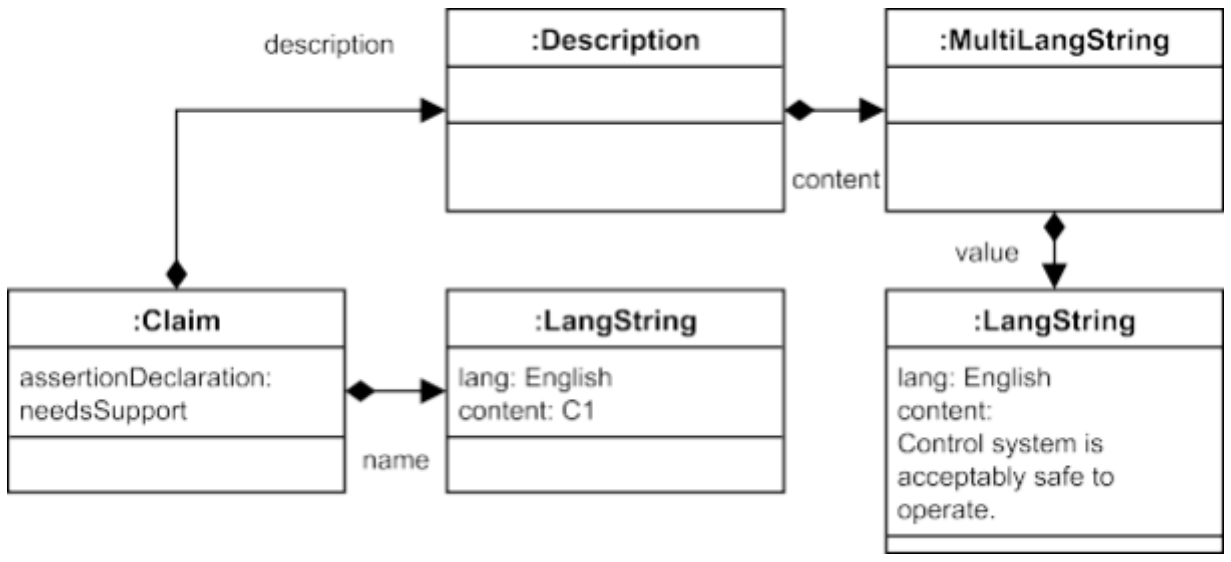

Figure 21: A **needsSupport** *Claim* in SACM.

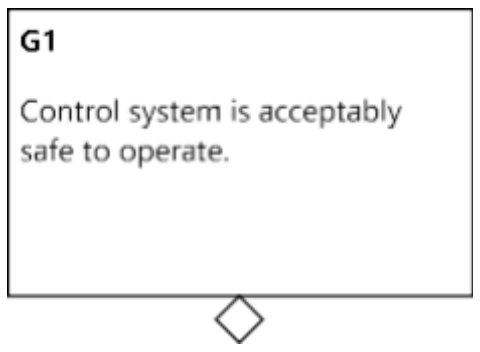

Figure 22: A undeveloped *Goal* in GSN.

A *Claim* marked as **assumed** is used to state an assumption in the argumentation. Figure 23 illustrates an **assumed** *Claim* in SACM (where the user assumes that *all hazards have been identified* ). The users are responsible to declare that a *Claim* is **assumed** when they make assumptions about the system. The equivalence for **assumed** *Claim* is an *Assumption* in GSN, as shown in Figure 24.

[16]Details of subsequent *Goal* s are omitted.

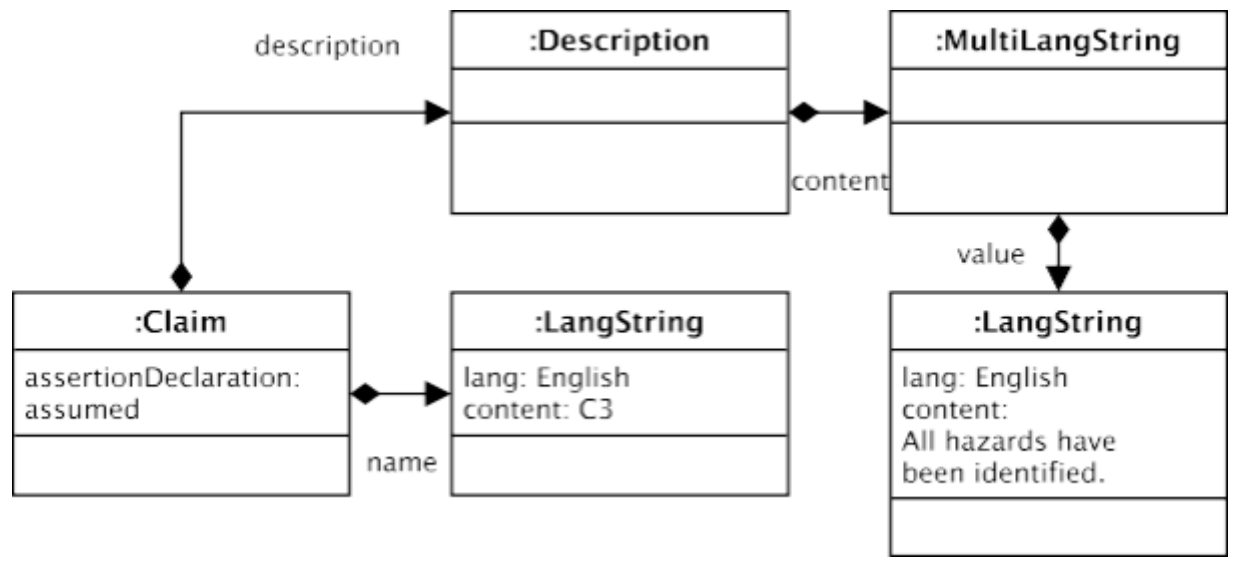

Figure 23: An **assumed** *Claim* in SACM.

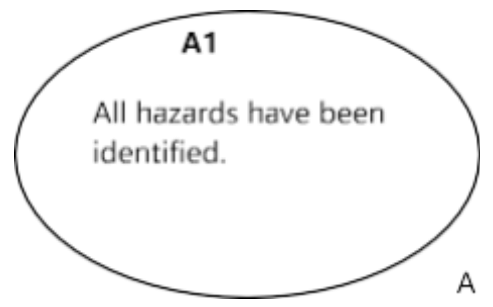

Figure 24: An *Assumption* in GSN.

A *Claim* marked as **axiomatic** is used to state a well agreed fact (presumably among stakeholders), which does not need any further support by arguments/evidence. The equivalent of an **axiomatic** *Claim* in GSN is a *Justification*, which does not need further supporting argument/evidence to support its content. Figure 25 and Figure 26 illustrates the use of **axiomatic** *Claim* and *Justification* respectively.

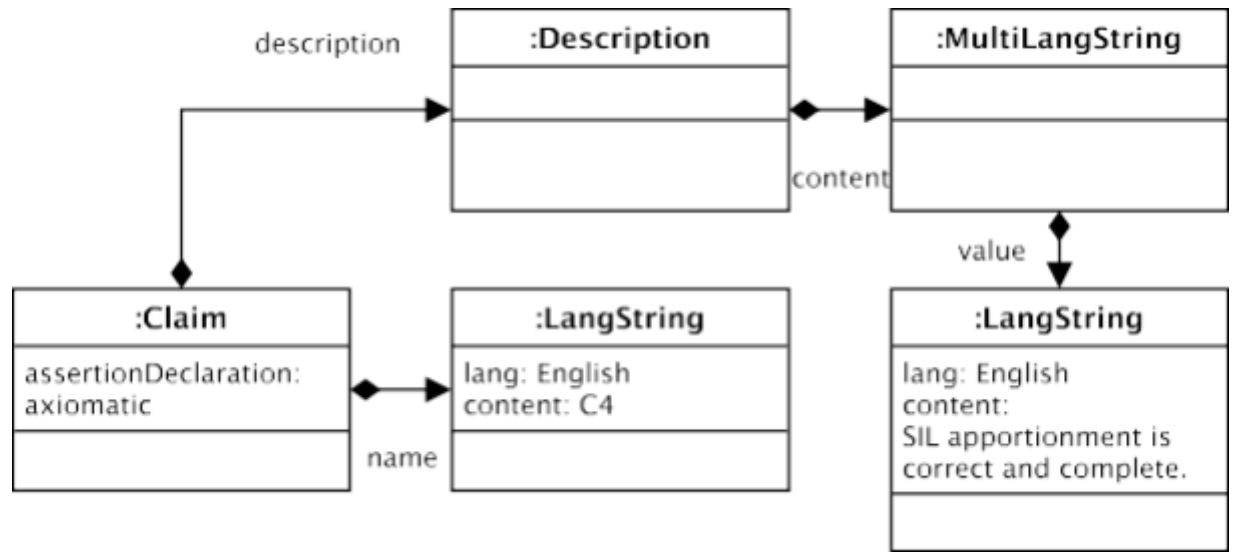

Figure 25: An **axiomatic** *Claim* in SACM.

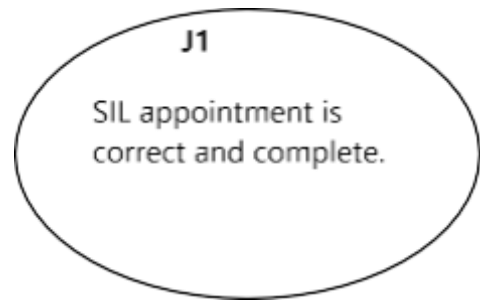

Figure 26: A *Justification* in GSN.

A *Claim* is marked as *defeated* if the truth of the claim is proven to be false by counter argument/evidence. Figure 27 provides an example of a **defeated** *Claim*, the *Claim* C1 is connected with another *Claim* (we do not consider the detail of this *Claim* in this example) by an *AssertedInference*, but with its *+isCounter* feature to be *true*. This means that the *AssertedInference* is a counter-argument, which negates the truth of *Claim*, if the *+source* of the *AssertedInference* is *true*. There is no equivalence for *defeated Claims* in GSN. As previously discussed, the users of SACM are able to selectively disclose the content of an *ArgumentPackage* via the use of *ArgumentInterface*. Figure 28 illustrates how an *ArgumentPackageInterface* can be used. In this example, a *Claim* C1 is held within an *ArgumentPackage*, which contains an *ArgumentPackageInterface* that in turn contains another *Claim*, which is a citation (its *+isCitation* is true). The *Claim* ‘cites’ C1 in the *ArgumentPackage* via the *+citedElement* feature (which is defined in the *SACMElement* element in the

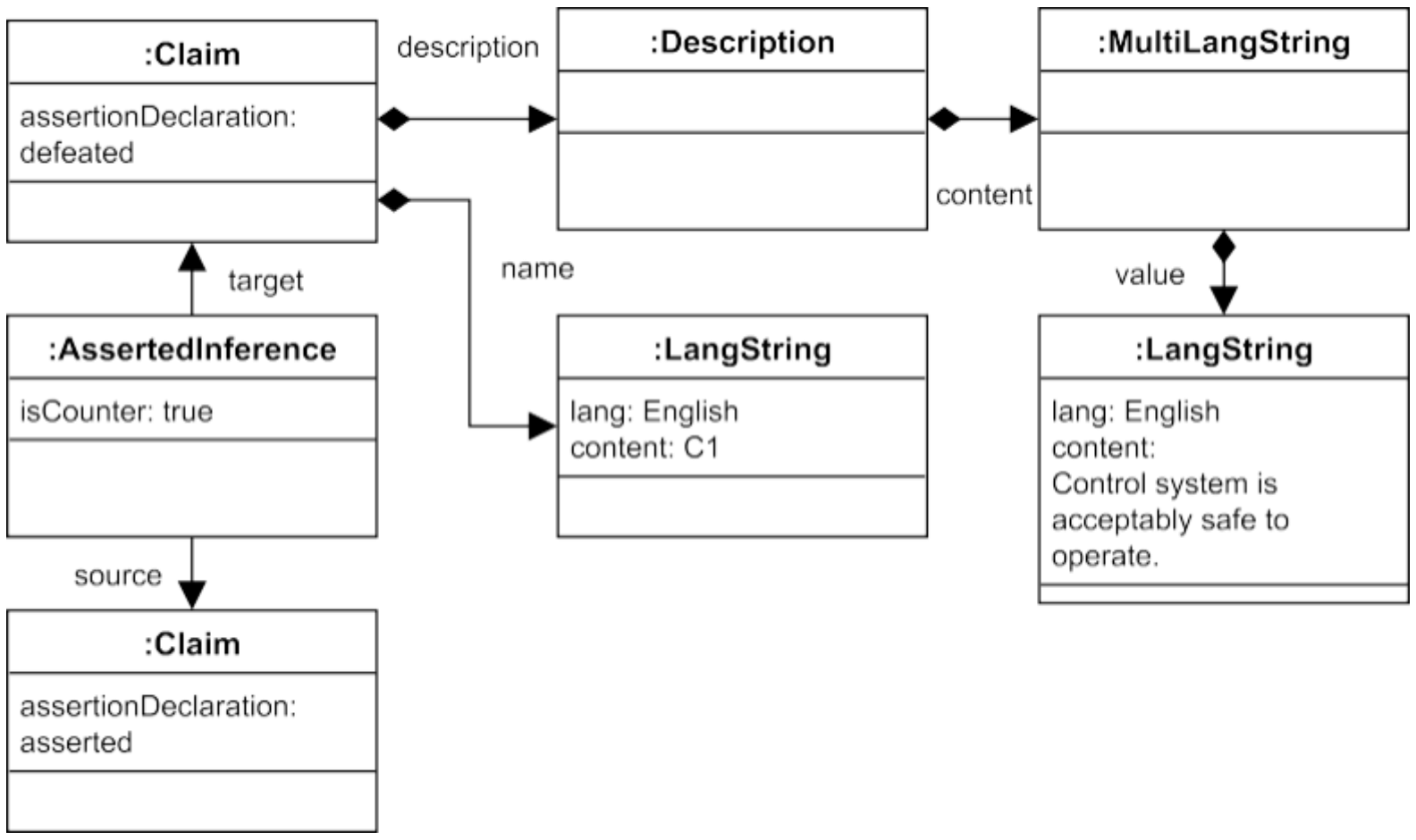

Figure 27: A **defeated** *Claim*.

***Base*** package). Hence, the ***asCited*** declaration on the ***Claim*** should be used, which means that the truth of the ***Claim*** depends on the ***Claim*** that it cites (in this case, ***Claim*** C1). There is no equivalence of ***cited Claim*** in GSN. The

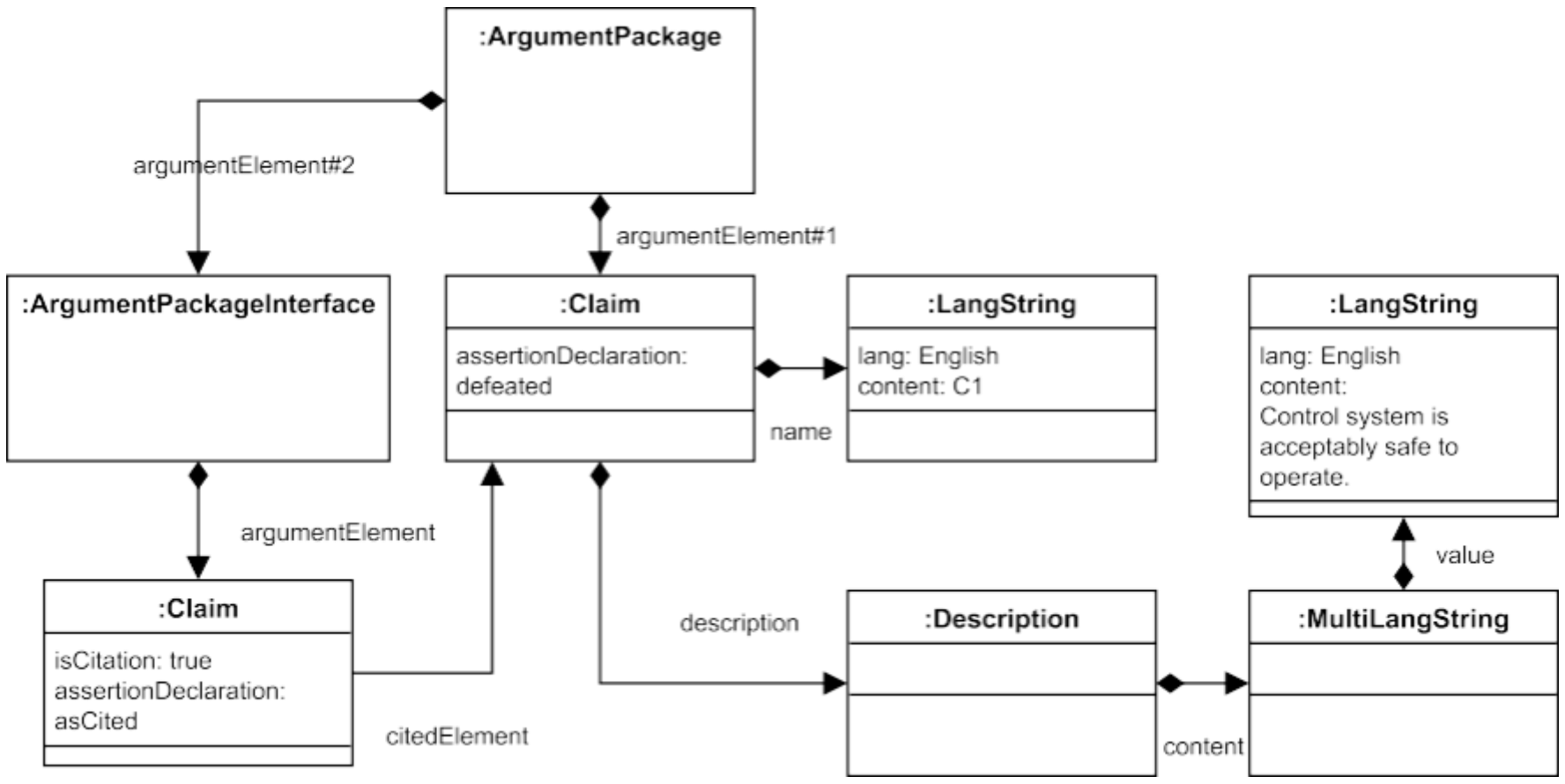

Figure 28: An **asCited** *Claim*.

***+isCitation*** and ***+citedElement*** features can be used in the same way in ***Artifact-PackageInterface*** and ***TerminologyInterface***, should the developers of assurance cases wish to disclose information in their corresponding packages.

### 5.2. *Example: AssertedRelationships and ArgumentReasoning*

The intent of *Claim*s is distinguished using *AssertionDeclaration*s. Different types of *Claim*s are connected using different *AssertedRelationship*s to form a structured argumentation.

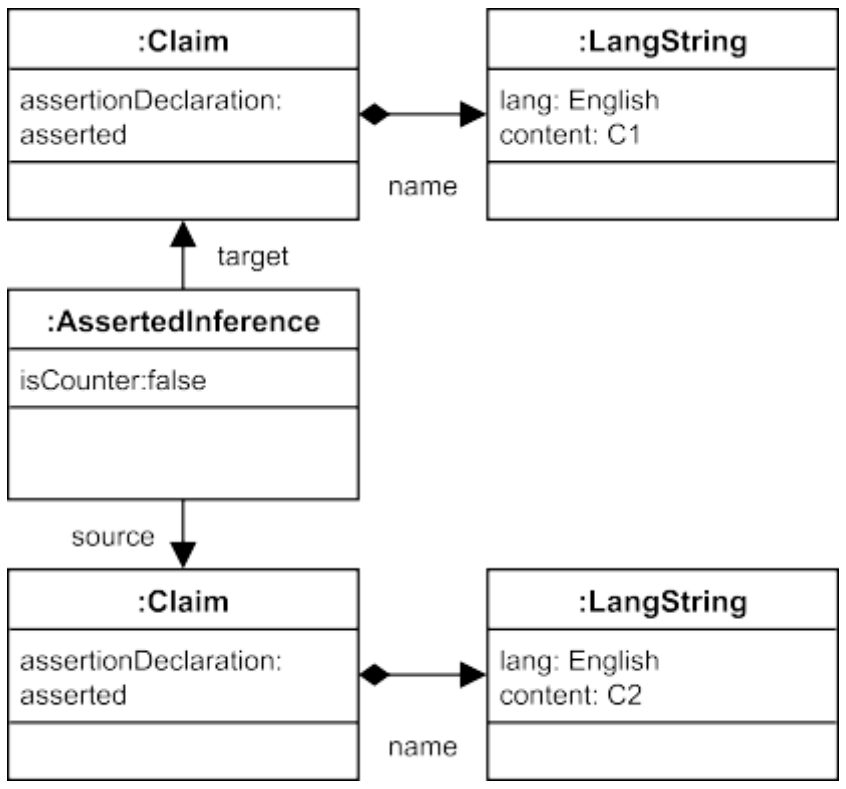

Figure 29: An example of **AssertedInference**.

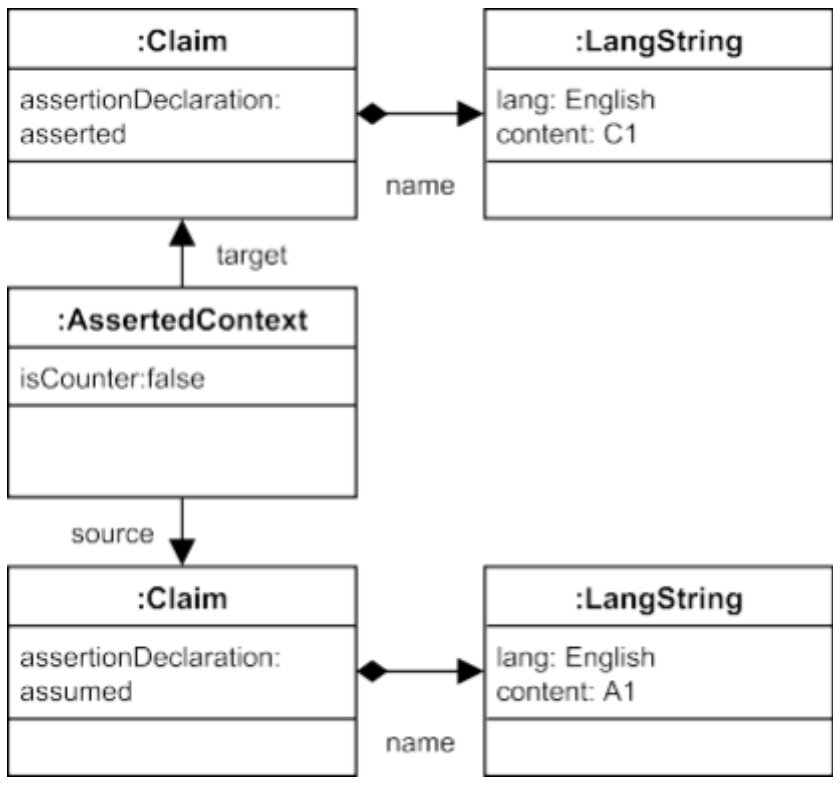

Figure 30: An example of **AssertedContext**.

The *AssertedInference* denotes the inference between one or more *Assertion*s and another *Assertion*. Figure 29 provides an example, where the truth of *Claim* C1[17] is inferred from the truth of Claim C2.

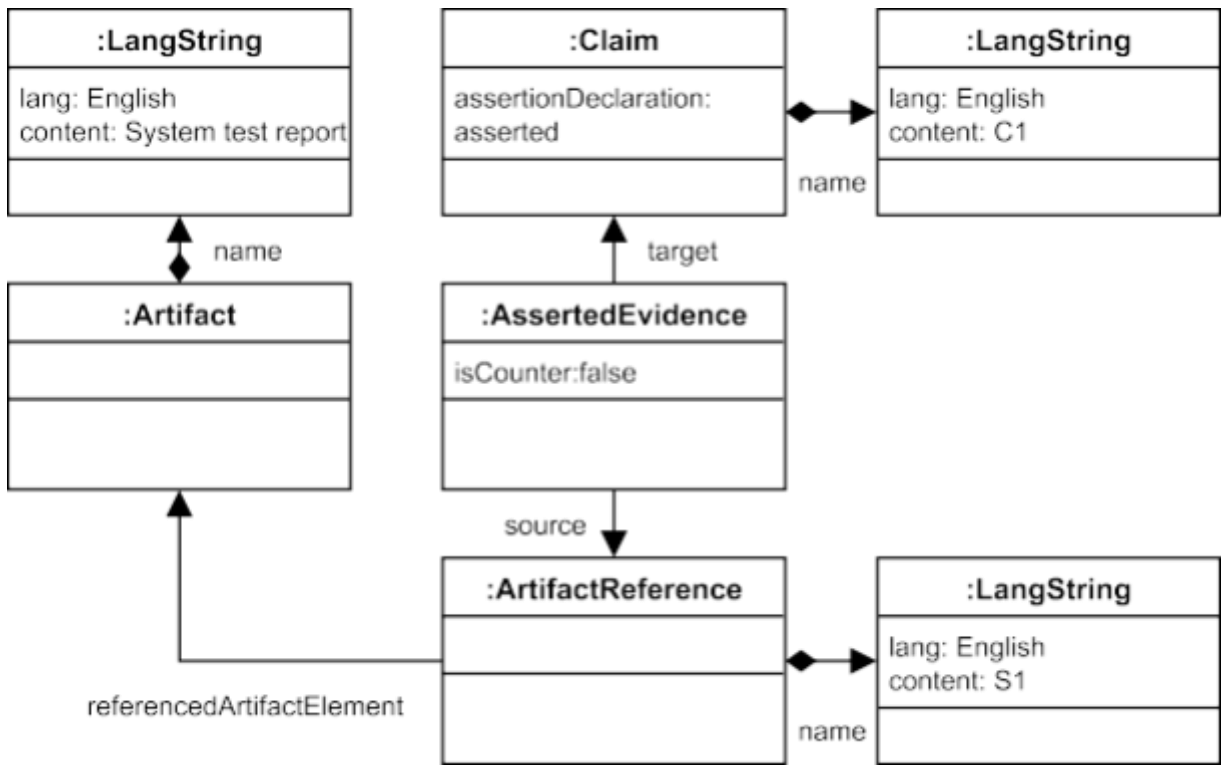

Figure 31: An example of **AssertedEvidence**.

The *AssertedContext* connects contextual *Assertion*s to an *Assertion*. Figure 30 provides an example, where the *Claim* A1 provides contextual informa-tion for ***Claim*** C1.

[17]Description details are omitted to make the model clearer.

The ***AssertedEvidence*** connects evidence to an ***Assertion***, Figure 31 provides an example, where the ***ArtifactReference*** S1 (which refers to an ***Artifact*** organised in an ***ArtifactPackage***, details omitted) provides evidence for the ***Claim*** C1.

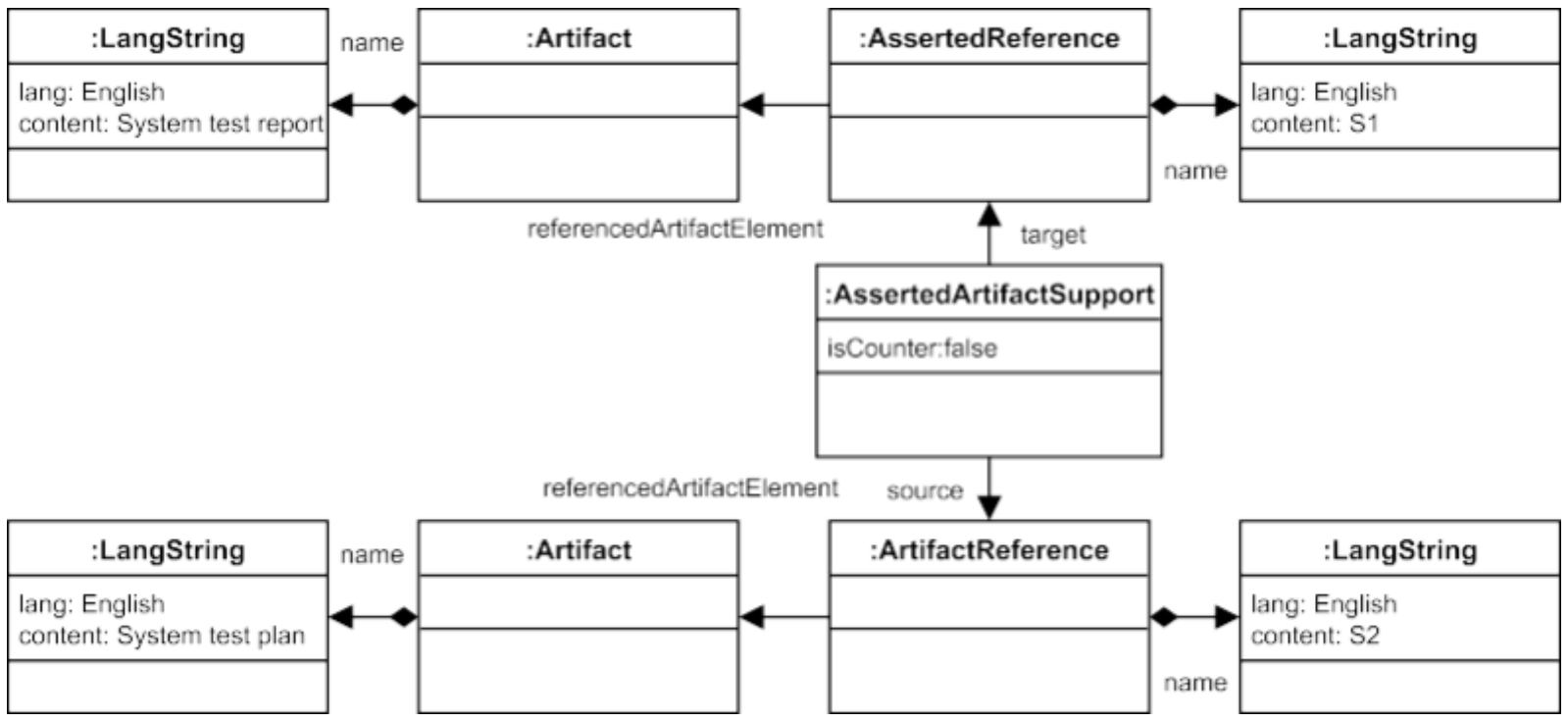

Figure 32: An example of **AssertedArtifactSupport**.

The ***AssertedArtifactSupport*** connects supporting artifacts to an artifact. Figure 32 provides an example, where ***ArtifactReference*** S2 (which cites an ***Artifact System test plan***) supports the ***ArtifactReference*** S1 (which cites the ***Artifact System test report***), in the sense that the system test plan supports the system text report. The ***AssertedArtifactContext*** ~~is also used on ***ArtifactReferences***, except that~~ it connects contextual artifacts to another artifact.

As previously mentioned, all elements that extend ***ArtifactElement*** can be considered as an artifact. There are ***supporting*** and ***contextual*** relationships between elements rather than those in the ***Artefict*** package. For example, an assurance case developer may want to express that one ***ArgumentPackage*** bears supporting arguments to another ***ArgumentPackage***, in which case the user may want to use ***AssertedArtifactSupport*** to connect two ***ArtifactReference*** which reference these two ***ArgumentPackage***s.

Sometimes, a reason of ***AssertedRelationship***s can be attached to further clarify the reasons of the ***AssertedRelationship***. In this case,***ArgumentReasoning*** is used for this purpose. Figure 33 illustrates the use of ***ArgumentReasoning***, where top level ***Claim*** C1 is backed by its sub-***Claim*** C2, which are connected by an ***AssertedInference***. The user may choose to give the ***AssertedInference*** a reason with the use of ***ArgumentReasoning***. In this case, an ***ArgumentReasoning*** S1 is attached to the ***AssertedInference***, which states that the argument is made by ***Argument over all identified hazards***. Supportive and contextual information can also be associated to ***ArgumentReasoning*** via the use of ***AssertedSupport*** and ***AssertedContext***. Since an ***ArgumentReasoning*** is not an ***Assertion***, there is little value in arguing the soundness of the ***ArgumentReasoning***.

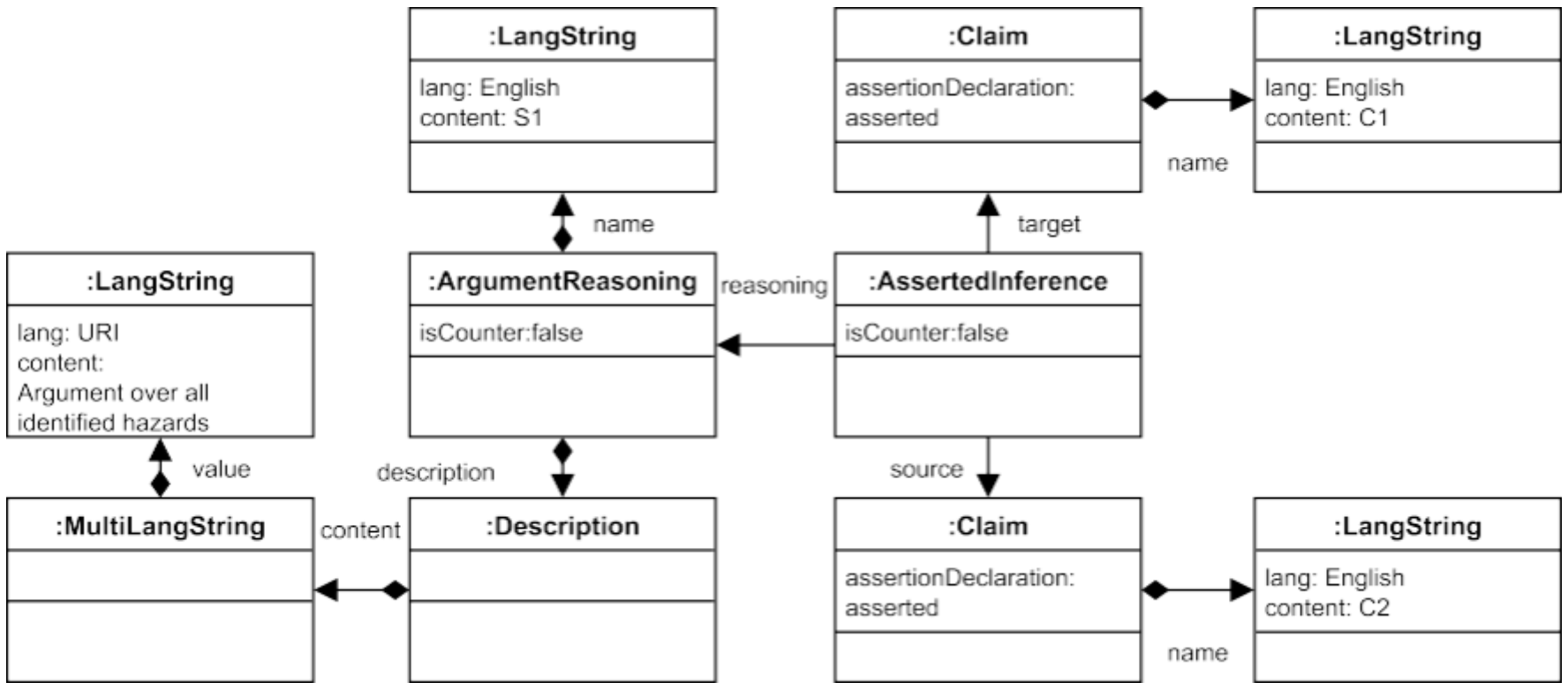

Figure 33: An example of **ArgumentReasoning**.

### 5.3. *Example: Argumentation Patterns*

In GSN, there is a concept of ***GSN Pattern***s [19], which are essentially successful GSN safety case templates that can be re-used. SACM provides similar concepts so that the user can construct patterns in SACM, and at a later stage, to ***instantiate*** the patterns by populating actual system information in the patterns. Figure 34 shows an argumentation pattern in SACM. For comparison, the content of the pattern is identical to the GSN pattern shown in Figure 8. Note that some details are omitted due to the complexity of the SACM model. There is a ***TerminologyPackage*** in the upper part of Figure 34 named TP1, within which contains an ***Expression*** and a ***Term***[18]. In the lower part of the figure, there is an ***ArgumentPackage*** AP1, which contains the structured argumentation pattern (N.B. the packages are placed in this way to improve readability of the figure and does not imply the priority of the packages).

The top level ***Claim*** of this pattern is G1, which maps to the ***Goal*** G1 in Figure 8. Note that G1's ***+isAbstract*** is set to ***true*** since this is an abstract ***Claim*** (a template). Now we focus on the ***description*** of G1, within which an ***ExpressionLangString*** is used. The ***ExpressionLangString*** refers to the ***Expression*** in ***TerminologyPackage*** TP1 (which contains value: ***{System X} is safe***). The curly brackets are a convention in GSN, which are called ***role***s in GSN. ***Role***s are essentially place holders in the pattern, the contents enclosed in the curly brackets will be replaced by actual system information if the pattern gets instantiated. In this case, when the pattern is instantiated, {System X} will be replaced with the name(s) of actual system(s).

In the ***TerminologyPackage*** TP1, the ***Expression*** refers to the ***Term*** ***System X***. This is a typical use of ***Term*** in structured argumentation. When the pattern is ***instantiated***, the ***Term*** should have its ***+externalReference*** feature populated

[18]The containment is described in this way to improve comprehensibility.

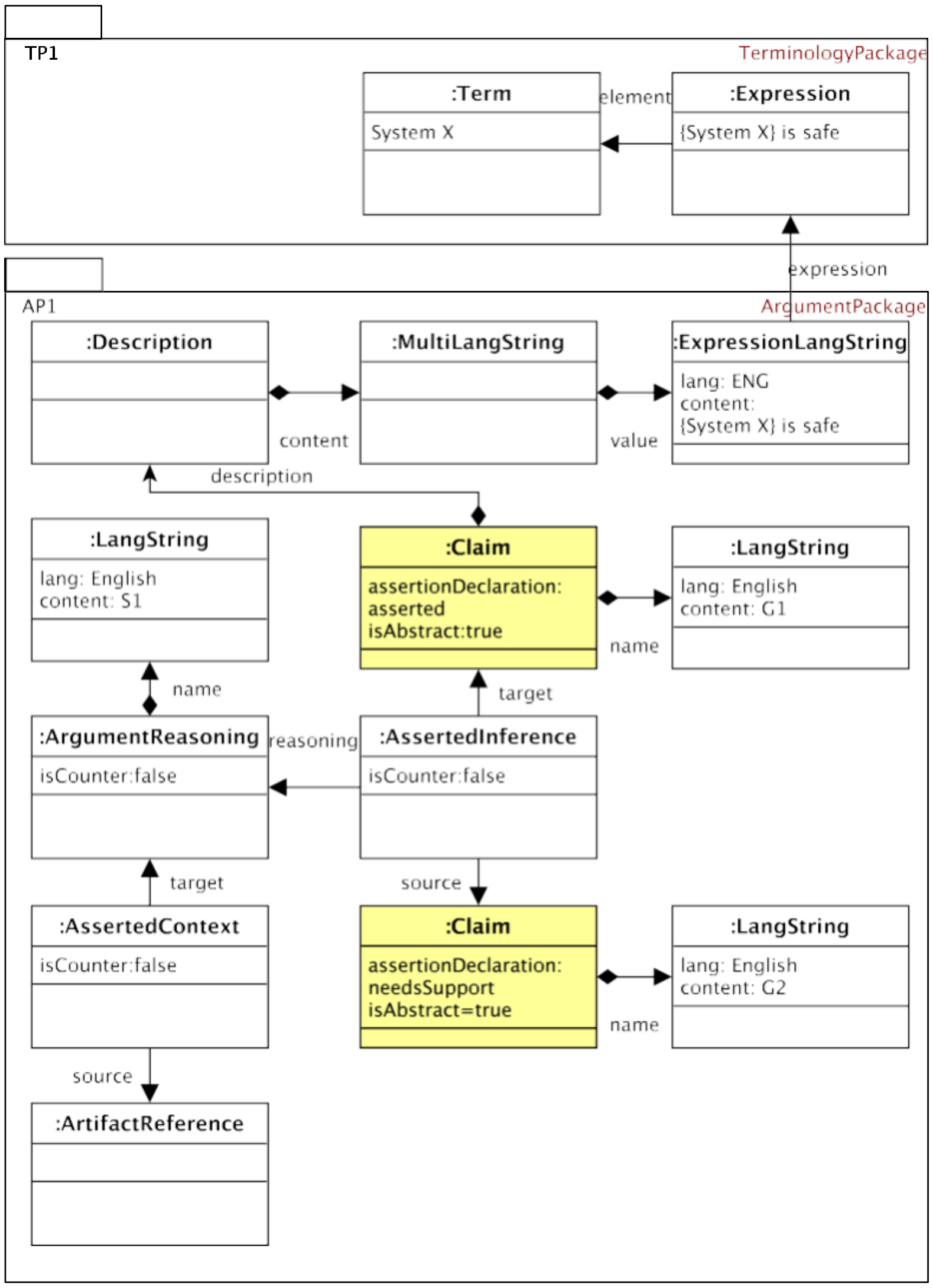

to import actual system information.

The structure of the argumentation in Figure 34 follows that in Figure 8, where the ***Claim*** G2 maps to ***Goal*** G2, and the ***ArgumentReasoning*** S1 maps to ***Strategy*** S1. Note that the ***Claim***s rendered in yellow are the templates, which have their ***+isAbstract*** feature set to true.

GSN pattern instantiation is often a manual procedure as safety case developers need to comprehend the GSN pattern and replace the ***role***s in the pattern with actual system information. In [12], a model-based approach is proposed, which makes use of a ***weaving model*** to link the elements in the GSN with elements in system models. This is typically due to the fact that in GSN there are no means to specify instantiation rules for GSN patterns. In SACM, as previously mentioned in Section 4.3, the element ***ImplementationConstraint*** can be used to specify the instantiation rules for patterns[19]. The users of SACM can associate ***ImplementationConstraint*** to any element and the use of language is not restricted. Evidently, pattern instantiation needs tool support and a model management engine to execute the ***ImplementationConstraint***s. The language used in ***ImplementationConstraint*** are not limited to computer languages. ***ImplementationConstraint*** s described in natural languages can also be used as instantiation rules, except that the instantiation procedure is limited to manual only.

When a SACM model instantiates a SACM pattern, it can relate to the SACM pattern via the ***+abstractForm*** feature. For example, if a ***Claim*** C1 created by instantiating the template G1, it can refer to G1 via the ***+abstractForm*** for future reference.

### 5.4. *Example: A case study on the European Train Control Systems (ETCS)*

As previously discussed, in SACM, assurance cases can be integrated to form an overall assurance case. Integrating assurance cases is a typical task performed when system components (or sub-systems) are integrated to form an overall system. Sometimes, system components/sub-systems are developed independently, together with their assurance cases. In safety-related domains, integration of assurance cases of components/sub-systems to argue the dependability of the to-be-integrated system is the premise of system integration.

To illustrate how assurance case integration is performed in SACM, we provide an example[20] taken from an engineering case study in European Train Control System (ETCS) we encountered in the DEIS project [32]. In this example, we consider the scenario where the assurance cases of on-board and side-track components of ETCS are integrated to form an overall assurance case.

The example SACM model for the ETCS case study is shown in Figure 35. For simplicity, we only show the top level ***Claim***s of the assurance cases. In Figure 35, there are three ***AssuranceCasePackage***s: On-Board ACP (at the

[19]Feature **F7** in Section 2.4.

[20]We reduced the complexity of the model structure by incorporating the name and description of the elements directly in the elements themselves.

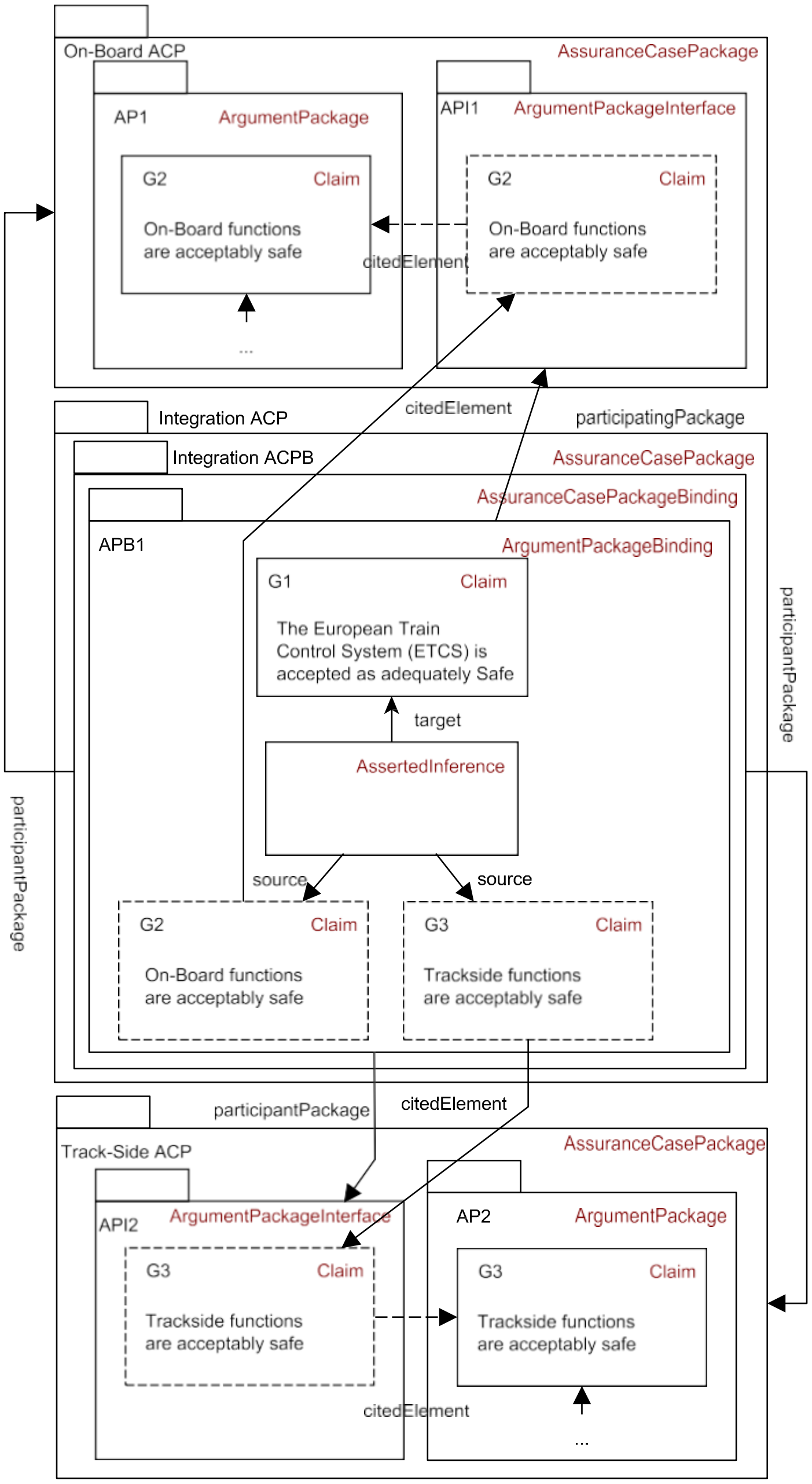
On-Board ACP
AssuranceCasePackage
AP1
ArgumentPackage
API1
ArgumentPackageInterface
G2
Claim
On-Board functions are acceptably safe
G2
Claim
On-Board functions are acceptably safe
citedElement
...
citedElement
participatingPackage
Integration ACP
Integration ACPB
AssuranceCasePackage
AssuranceCasePackageBinding
APB1
ArgumentPackageBinding
G1
Claim
The European Train Control System (ETCS) is accepted as adequately Safe
target
AssertedInference
source
source
G2
Claim
On-Board functions are acceptably safe
G3
Claim
Trackside functions are acceptably safe
participantPackage
participantPackage
participantPackage
citedElement
Track-Side ACP
AssuranceCasePackage
API2
ArgumentPackageInterface
AP2
ArgumentPackage
G3
Claim
Trackside functions are acceptably safe
G3
Claim
Trackside functions are acceptably safe
citedElement
...

top of the figure) is the assurance case for the on-board component of ETCS; Track-Side ACP (at the bottom of the figure) is the assurance case for the side-track component of ETCS; and Integration ACP (in the middle of the figure) is the integrated assurance cases which integrates the two component assurance cases together.

For On-Board ACP, ***ArgumentPackage*** AP1 contains the argument regarding the safety of the on-board component. As discussed in Section 4.2, system engineers may wish only to disclose the top level ***Claim*** externally, hence ***ArgumentPackageInterface*** AP1 is used which contains a citation of G2 which will be referenced externally. Same principle is applied to Track-Side ACP, where the top level ***Claim*** G3 is cited in the ***ArgumentPackageInterface*** API2. It is to be noted that both On-Board ACP and Track-Side ACP contains their ***ArtifactPackage***s and ***TerminologyPackage***s, which are not shown due to the complexity of the model structure.

To integrate On-Board ACP and Track-Side ACP, ***AssuranceCasePackage*** named Integration ACP is created. Integration ACP contains an ***AssuranceCasePackageBinding*** (Integration ACPB) specifically to bind On-Board ACP with Track-Side ACP. Within Integration ACPB, an ***ArgumentPackageBinding*** (APB1) is used to bind API1 and API2 via the +***participantPackage*** feature. In APB1, top level ***Claim*** G1 argues the safety of ETCS and two supporting ***Claim***s G2 and G3 are in place. Note that G2 and G3 are citation ***Claim***s which cites G2 in API1 (which in turn cites G2 in AP1) and G3 in API2 (which in turn cites G3 in AP2). It is also to be noted that within the ***AssuranceCasePackageBinding*** (Integration ACPB), there are also ***ArtifactPackageBinding***s and ***TerminologyPackageBinding***s which binds the artifacts and expressions used in the on-board component and track side component.

The integration of assurance cases in SACM is achieved using various package bindings. It is also possible to include additional arguments in the binding ***AssuranceCasePackage*** when deemed necessary. Users of SACM may also argue the trustworthiness of the cited ***Claim***s in other packages to ensure the confidence in citing argument elements.

This simplified example illustrates that the users of SACM are able to integrate assurance cases using the facilities provided. It is to be noted that assurance case composition is a complex task, for there may be subtle dependencies among the systems. Modular assurance case construction makes no assumption that the safety/security of the whole system is guaranteed by a composition of arguments about the safety/security of the parts. It remains the responsibility of the assurance case developers to ensure that each assurance case module correctly identifies its dependencies on other assurance cases in order to assure the composed system.

## 6. Metamodel for existing notations and the transformations to SACM

SACM is designed to support existing safety notations such as GSN and CAE. In previous sections, we briefly demonstrated the usage of SACM elements by comparing them with GSN notations. In this section, we provide a GSN metamodel and a CAE metamodel that are compliant to SACM. We also discuss the transformation from GSN and CAE to SACM.

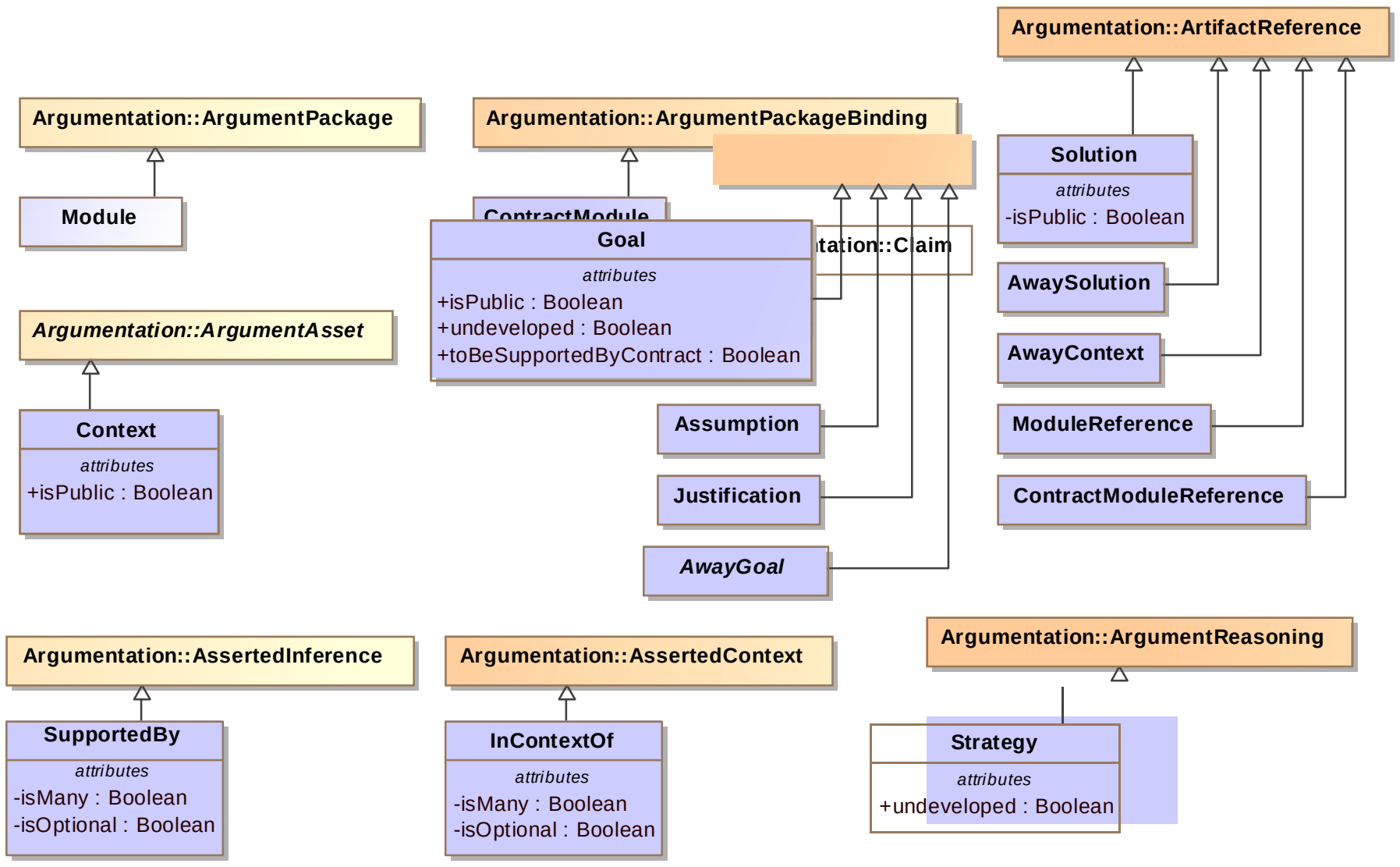

Figure 36: SACM compliant GSN metamodel.

### 6.1. *The GSN Metamodel and the inteoperability from GSN to SACM*

As previously discussed, SACM provides a richer set of features compared to GSN, which includes the ability to standardise evidential and informational artifacts in the models, the ability to standardise controlled vocabularies (expressions and terminologies), as well as modular organisation and integration of artifacts and terminologies. In general, creating a metamodel for GSN is a simple task, for there are only a few concepts that GSN captures. However, it is ideal to create the GSN metamodel by extending SACM elements, so that not only can the GSN metamodel inherit features provided by SACM, but it also makes the interoperability from GSN to SACM easier.

Our version of the GSN metamodel is shown in Figure 36[21]. In GSN, argumentations are organised in ***Module***s, which is made as a sub-type of ***ArgumentPackage*** in SACM; ***ContractModule*** is essentially contract that binds ***Module***s together, thus it is a sub-type of ***ArgumentPackageBinding***.

Elements ***Goal***, ***Assumption***, ***Justification*** and ***AwayGoal*** are made sub-types of ***Claim*** in SACM. A ***Goal*** can be ***uninstantiated***, which basically means it is abstract, this is captured by the ***+isAbstract*** feature in ***SACMElement*** class of SACM. A ***Goal*** can be ***public***, which is deprecated in SACM, a ***Goal*** can also be ***undeveloped*** and ***toBeSupportedByContract***, which are captured individually.

[21]All GSN elements are rendered in blu

Elements ***Solution***, ***AwaySolution***, ***AwayContext***, ***ModuleReference*** and ***ContractModuleReference*** are sub-types of ***ArtifactReference*** in SACM as they refer to artifacts that contain information they represent. ***Context*** is a slightly complicated concept, as it can either be a statement stating the context of a ***Claim***, or it can refer to contextual information stored in an artifact. Thus, ***Context*** is made a sub-type of ***ArgumentAsset***.

***SupportedBy*** is made a sub-type of ***AssertedInference*** and ***InContextOf*** is made a sub-type of ***AssertedContext***. ***Strategy*** is made a sub-type of ***ArgumentReasoning*** for it explains the intention of an ***AssertedRelationship***.

The way that the GSN metamodel is created makes it inherently compatible with SACM, so that it can be used as an ***ArgumentPackage***, and be organised in an ***AssuranceCasePackage***, with its supporting artifacts organised in ***ArtifactPackage***s and controlled vocabularies organised in ***TerminologyPackage***s.

To enable interoperability from GSN to SACM, we provide a model-to-model transformation[22] defined using the Epsilon Transformation Language (ETL) [20]. Most of the transformation is straight forward - instances of the GSN elements are transformed into instances of the SACM elements that the GSN elements extend. There is one exception: the transformation from ***Strategy*** to ***ArgumentReasoning***. ***Strategy*** in GSN is a node in the structured argumentation, where ***ArgumentReasoning*** is a node associated to an ***AssertedRelationship*** (which is an edge). Hence, additional analysis is needed to perform the GSN2SACM transformation. This requires analysis to be performed during the transformation, which is shown in Algorithm 1.

Algorithm 1 defines the transformation rule ***Strategy2ArgumentReasoning***. Line 1 retrieves the ***Strategy*** to be transformed; Line 2 creates a new ***ArgumentReasoning*** ; Line 3 and 4 transform the ***name*** and ***description*** of the ***Strategy*** to the ***ArgumentReasoning*** (the equivalent() operation in ETL divert control to corresponding transformation rules that transform ***Name*** to ***Name***, and returns the transformed model element). Line 5 checks if the ***Strategy*** is uninstantiated, and then in line 6, make the ***ArgumentReasoning abstract***. Line 8 retrieves the incoming ***SupportedBy*** for the ***Strategy***. Line 9 retrieves all outgoing ***SupportedBy*** for the ***Strategy*** (In GSN, the flow of an argument goes from top to bottom. Thus, for a ***Strategy***, there is one incoming ***SupportedBy*** and one or many outgoing ***SupportedBy***s). Line 10 retrieves the ***Goal*** from the incoming ***SupportedBy***. Line 11 retrieves the ***Goal***s from the outgoing ***SupportedBy***s. Then, in line 13 an ***AssertedInference*** is created. It is worth noting that in SACM, the inference flows from bottom to top. Thus the source and target of the ***AssertedInference*** are the transformed counterparts of ***supportedByToGoals*** in line 11 and the transformed counterparts of ***supportedByFromGoal*** in line 10, respectively. It is to be noted that some developers use ***Solutions*** to directly support ***Strategy***. This is forbidden in the GSN standard (permitted SupportedBy con- nections are: ***Goal***-to-***Goal***, ***Goal***-to-***Strategy***, ***Goal***-to-***Solution*** and ***Strategy***-to- ***Goal*** ). This is one of the motivations why model-based assurance case is desired

[22] Available at: https://github.com/wrwei/MDERE/blob/master/technical

**Algorithm 1:** Transforming Rule Strategy2ArgumentReasoning

```
let strategy = Strategy to be transformed;
let argumentReasoning = new ArgumentReasoning;
argumentReasoning.name = strategy.name.equivalent();
argumentReasoning.description = strategy.description.equivalent();
if strategy.uninstantiated == true then
    argumentReasoning.isAbstract = true;
end
let incomingSupportedBy = the incoming SupportedBy to strategy;
let outgoingSupportedBys = all outgoing SupportedBys from strategy;
let supportedByFromGoal = Goal from incomingSupportedBy
let supportedByToGoals = all Goals from outgoingSupportedBys that connects
  to strategy;
if supportedByToGoal is not empty then
    let assertedInference = new AssertedInference;
    assertedInference.target = supportedByFromGoal.equivalent();
    for goal in supportedByToGoals do
        assertedInference.source.add(goal.equivalent());
    end
end
```

- automated model validation can be performed to check the well-formedness of assurance cases.

### *6.2. The CAE metamodel and the interoperability from CAE to SACM*

Claims-Arguments-Evidence (CAE) [4] is another widely used graphical notation for assurance case construction. Concepts in CAE are similar to those in GSN. There has not been an official metamodel defined for CAE. Thus, we provide our own version of CAE that extends SACM, shown in Figure 37[23].

Unlike GSN and SACM, CAE does not support modularity. Therefore, we introduced three new concepts in CAE, ***CAEModule***, ***CAEModuleInterface*** and ***CAEModuleBinding***, which extend ***ArgumentPackage***, ***ArgumentPackageInterface*** and ***ArgumentPackageBiding*** respectively. In CAE, there is a notion of ***Claim***, which is semantically identical to ***Claim*** in SACM. We therefore create a ***CAEClaim*** element that extends ***Claim*** in SACM. The reason for this redundancy is that we want the CAE metamodel to be non-invasive to SACM. The ***Argument*** elements in CAE provide a description of the argument approach, which is functionally equivalent to ***ArgumentReasoning***, thus it is made as a sub-type of ***ArgumentReasoning***. ***Evidence*** is made as a sub-type of ***ArtifactReference*** because it is a reference to evidential materials. In CAE, there is a notion of ***Assumption***, which is made as a sub-type of ***Claim***.

In CAE, there are three types of relationships, the ***IsEvidenceFor*** connects ***Evidence*** with ***Claim***s, thus is made a sub-type of ***AssertedEvidence***; the ***IsSub-***

[23] All CAE elements are rendered in blue

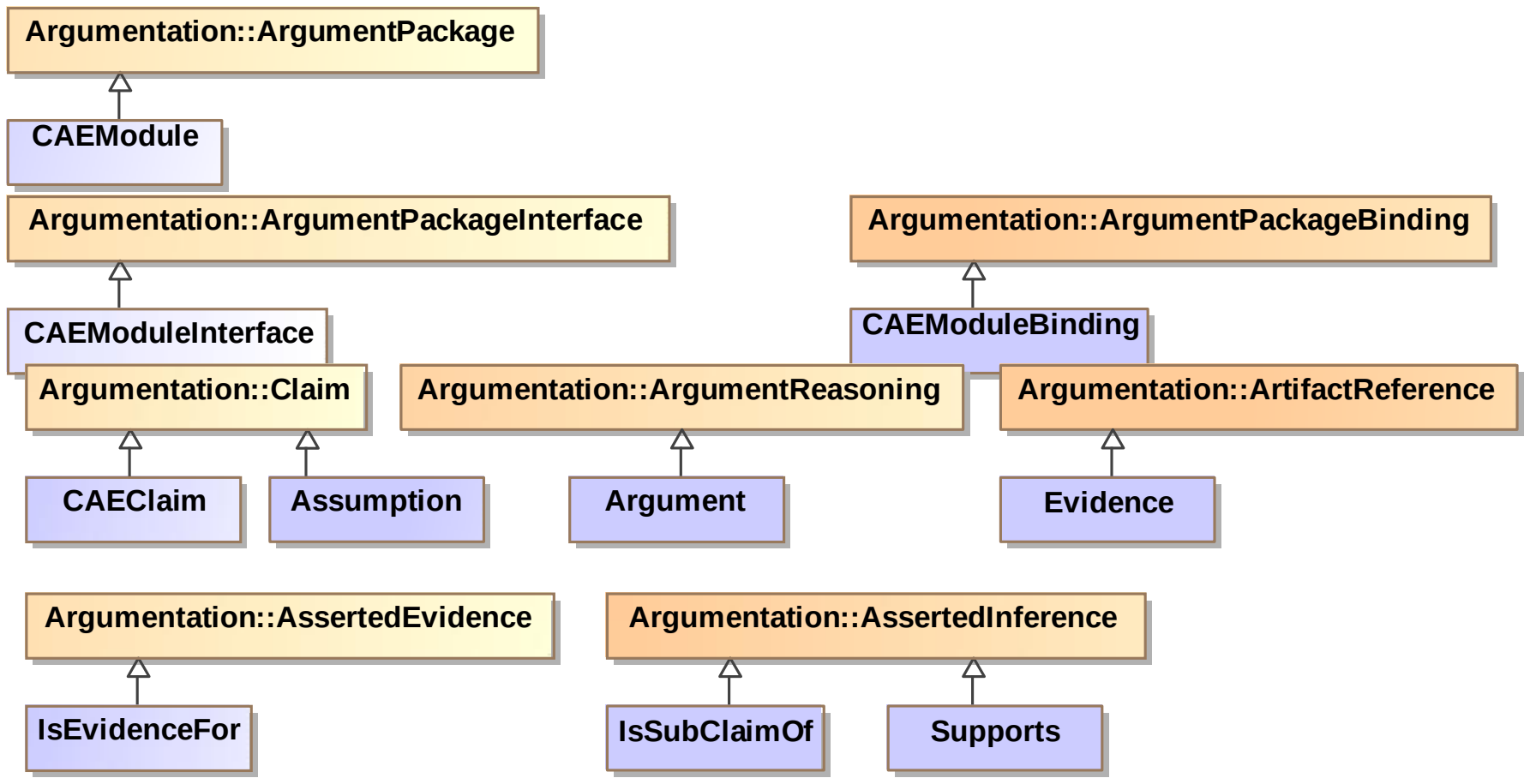

Figure 37: SACM compliant CAE metamodel.

*ClaimOf* relationship connects sub-*Claim*s to *Claim* and is made a sub-type of *AssertedInference*; the *Supports* relationship connects *Argument* s to *Claim* and is also made a sub-type of *AssertedInference*. Since there is no notion of modules in CAE, the argumentation is contained in *ArgumentPackage*s inherited from SACM.

The transformation from CAE to SACM is similar to the transformation from GSN to SACM (which is also implemented in Epsilon Transformation Language), with the same issue that *Argument* (which is a node in the structured argumentation) needs to be mapped to *ArgumentReasoning* (a node associated to an edge). The detailed transformation is made publicly available[24].

## 7. Tool Support and Future Work

### 7.1. Assurance Case Modelling Environment - ACME

With all its power in model-based system assurance, there is one drawback for SACM at the moment, which is the lack of concrete syntax, i.e. graphical representations of SACM[25]. Without graphical representations, it is typically difficult for engineers to construct SACM. Hence, in order to exploit the benefits provided by SACM whilst providing supports for existing assurance case approaches, we started developing a tool (Assurance Case Modelling Environment, ACME[26]) based on SACM and the GSN metamodel we discussed in

[24]Available at: https://github.com/wrwei/MDERE/blob/master/technical
[25]Graphical syntax for SACM is being developed by the authors.
[26]Available upon request.

Section 6.

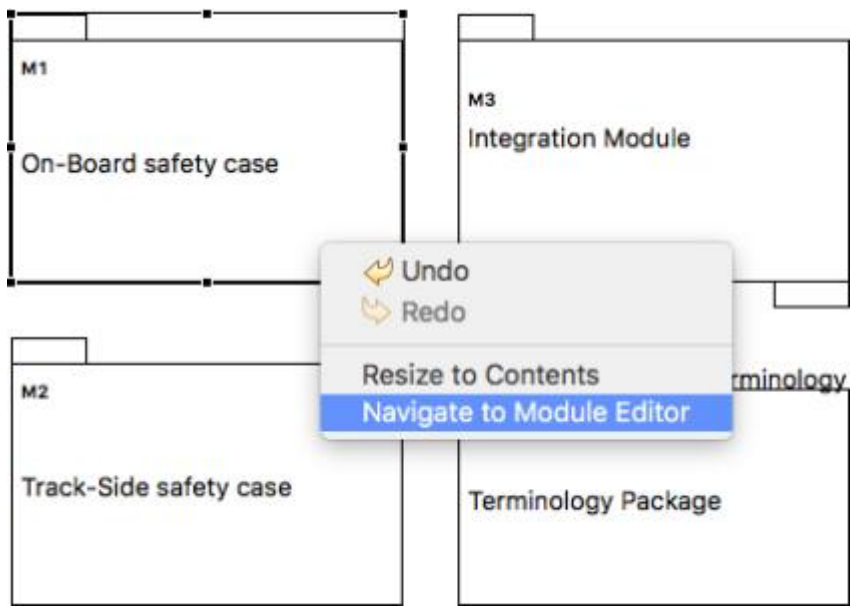

Figure 38: The Assurance Case Package View of ACME.

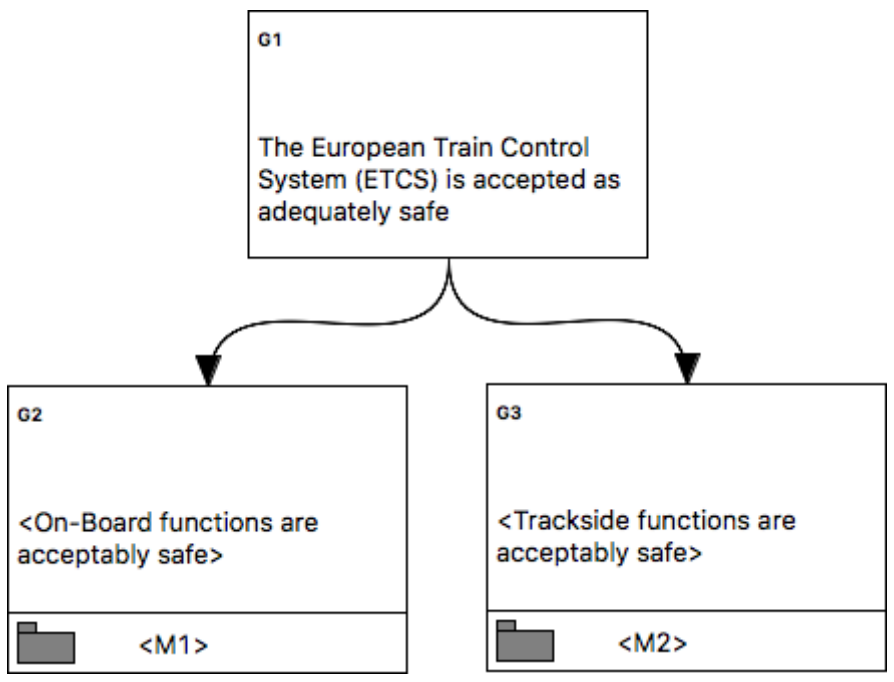

Figure 39: The Module View of ACME.

ACME is implemented using the Graphical Modelling Framework (GMF), which supports the creation of editors based on metamodels defined using the Ecore metamodel provided by the Eclipse Modelling Framework (EMF) [29]. As previously discussed, we created metamodels for GSN and CAE. The approach we take for ACME is to support GSN for the argumentation part of the assurance case (since there is no graphical syntax for the ***Argumentation*** component of SACM). In this sense, the users of ACME are able to create an assurance case using SACM, use GSN for the arguments, and then use SACM’ s ***Artifact*** and ***Terminology*** components for evidence-artifact traceability and controlled vocabulary. In this way, system assurance case practitioners are able to make the transition from GSN/CAE to SACM, from (mostly) non-model-based approach to uniformed model-based approach.

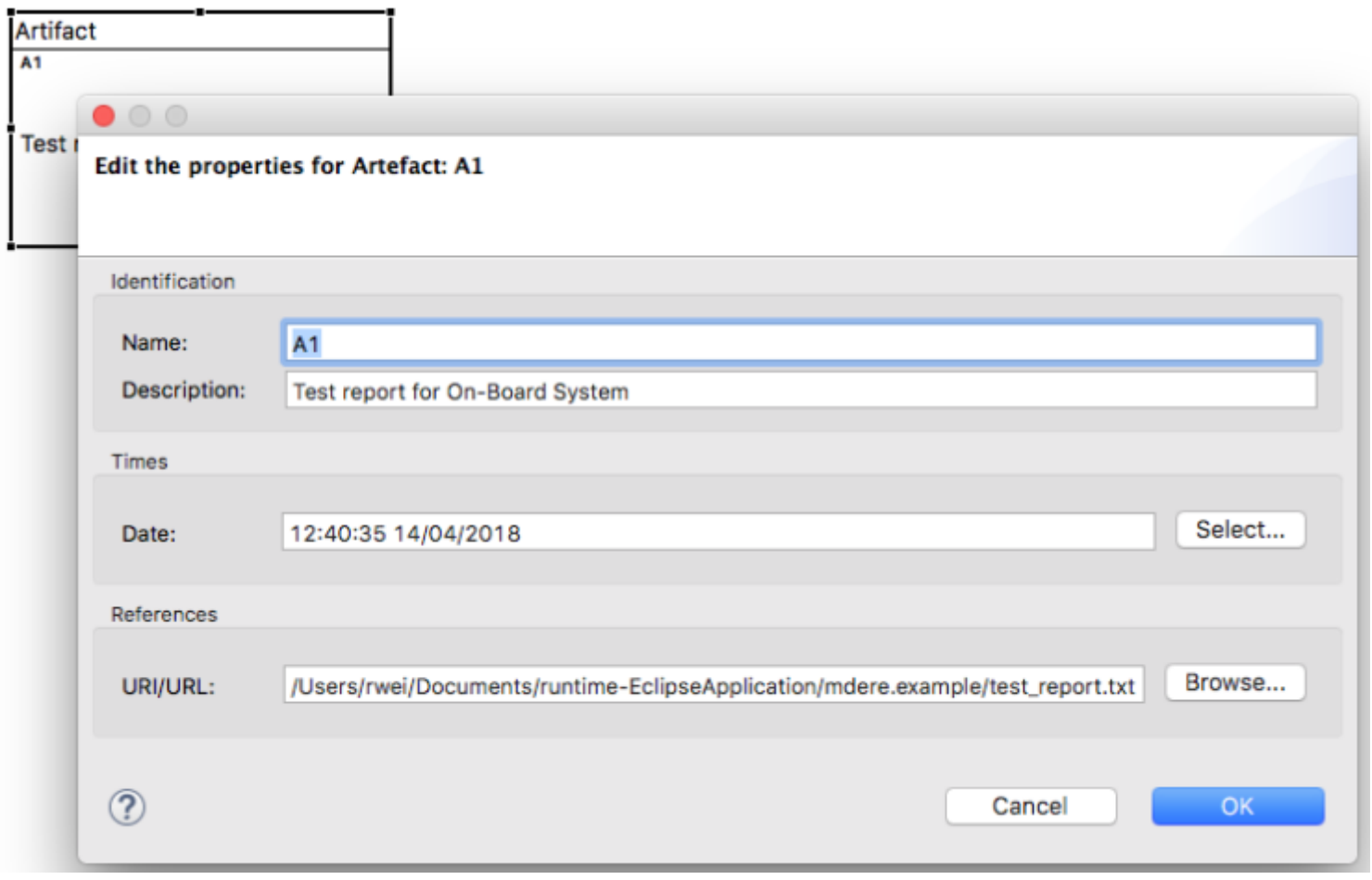

Figure 40: Creation of Artifact.

Figure 38 shows the Assurance Case Package View in ACME, within which the users are able to create Modules and Contract Modules of GSN, as well as elements defined in the *AssuranceCase* component of SACM. Figure 39 shows the Module view of ACME, where the users are able to create GSN elements.

The users are also able to create elements in the *Artifact* component of SACM (i.e. *Artifact*, *Activity*, *Event*, *Participant*, *Technique*, *Resource*, *Property*, and *ArtifactAssetRelationship*), Figure 40 demonstrates how an *Artifact* element can be created/edited. Creation tools are also provided for the elements in the *Terminology* component, where a *TerminologyPackage* is represented as a table in ACME. Figure 41 demonstrates how a *TerminologyPackage* and its contents can be created/edited.

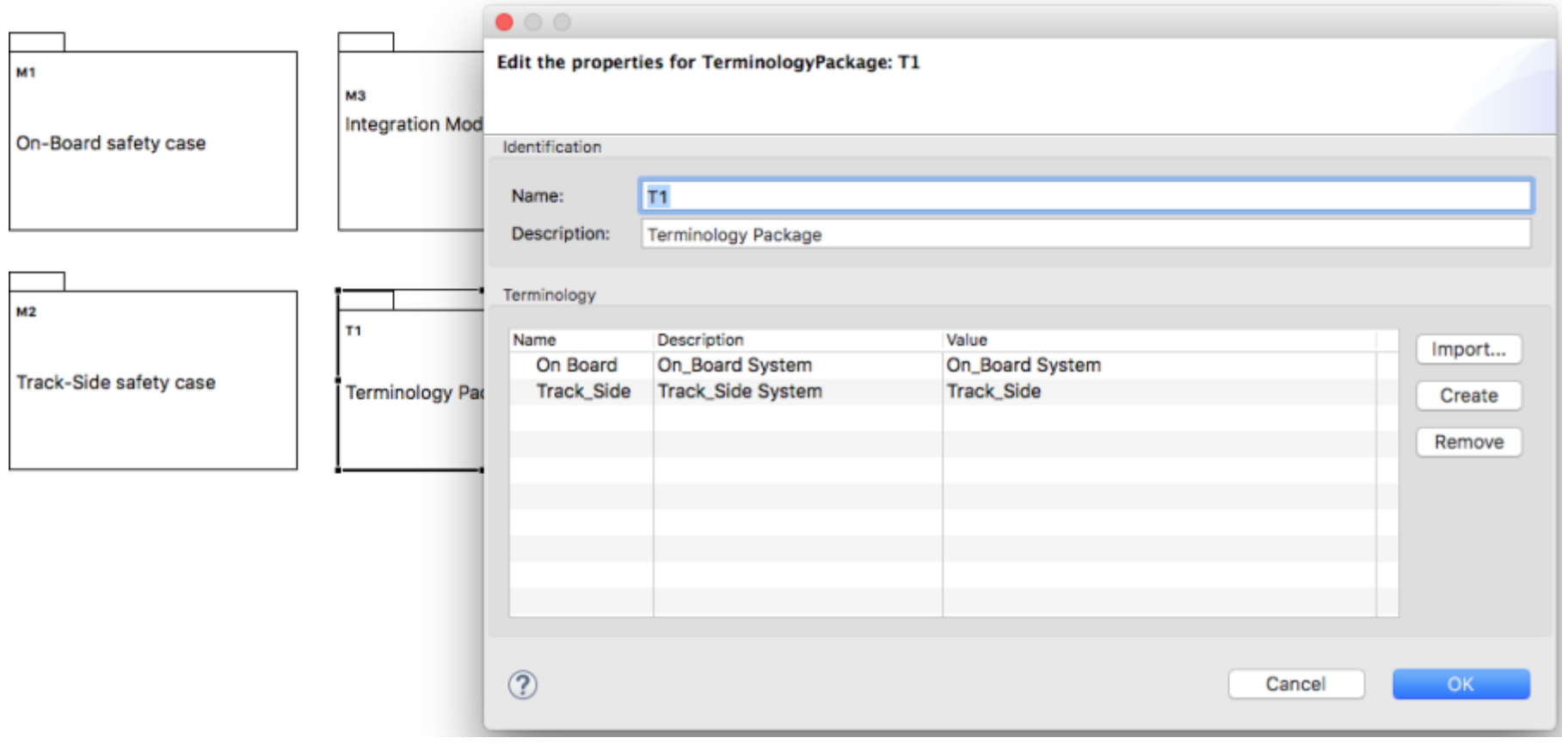

Figure 41: Editing a *TerminologyPackage*.

The idea behind ACME is to provide a transition for practitioners from GSN (and/or CAE[27]) to SACM, before the OMG standardises the graphical syntax of SACM. In this way, assurance case practitioners can continue to use GSN, whilst exploiting the features provided by SACM (explained in Section 2.4). When the graphical syntax of SACM is standardised[28], ACME is ready and will be the first tool to adopt the graphical syntax.

The users of ACME are also able to transform their GSN models to SACM, based on the model-to-model transformation from GSN to ACME. ACME integrates with the Epsilon platform, so that model management operations are supported on GSN/SACM models. A first step towards this direction is the support for GSN2SACM transformation, which is provided by ACME.

ACME is the first step towards an integrated modelling environment for SACM. ACME provides a transitional solution to model-based system assurance, in the sense that it enables existing assurance case approaches to be used in conjunction with SACM, in order to exploit the features provided by SACM such as evidence-artifact traceability, controlled vocabularies and multiple language support. ACME as it is at the moment, illustrates what can be done with SACM and model-based assurance cases created with SACM. However, it does not guarantee that assurance cases provided by ACME are compelling ones. It is still the responsibility of assurance case developers to develop accurate and correct assurance cases.

[27]CAE editor for ACME is under development

[28]As we are aware, this standardisation is in process

### *7.2. Future Work*

The next work in line for SACM is the standardisation of its graphical notations. As explained, SACM is not a metamodel for GSN and CAE, it is an independent metamodel for creating assurance cases using its own Argumentation component (which can be seen as the superset of GSN and CAE). Apart from modelling structured argumentataions, SACM can be used to to model evidence and the processes that manage the evidence (the Artifact component), and to model controlled vocabulary used in assurance cases (the Terminology component). Therefore, for engineers to create SACM models efficiently, it is necessary to introduce graphical notations so that tools like ACME can be built to enable the creation of SACM models.

With regard to future work for ACME, we aim at achieving the following objectives:

- Support for CAE. We aim to create the editor for CAE and integrate it to ACME.
- Support for legacy assurance cases. We aim to develop a set of model extraction mechanisms to extract information from assurance cases provided by existing tools and convert them to models that conform to our GSN/CAE metamodels.
- Model management tools. We aim to develop a set of model validation rules to check the well-formedness of GSN/CAE/SACM models. We also aim to develop a set of model-to-text transformation rules for assurance case report generation.
- Support for automated pattern instantiation. We integrate the Epsilon platform runtime[29] in ACME for various model management operations. We aim to explore how ***ImplementationConstraint***s can be utilised to hold Epsilon programs for pattern instantiation.
- And finally, support concrete syntax for SACM. We are contributing to the development of graphical notations for SACM at the moment, and will add the support for the notations in ACME once they are developed and evaluated.

[29] https://www.eclipse.org/epsilon/

## 8. Conclusion

In this paper, we identified the importance of model-based system assurance cases, for that they enable high level model management operations to be performed, and its potential applications in Open Adaptive Systems assurance at runtime. In the prespective of of modelling languages, SACM is more powerful than existing system assurance approaches (such as GSN and CAE), for its additional features listed (non-exhaustively) below;

- Fine grained modularity, for component based system assurance, as well as assurance case integration;
- Multiple language support, to support multiple natural languages, as well as computer languages;
- Controlled vocabulary, to standardise terminologies used in assurance cases;
- Ability to argue the trustworthiness of arguments, so that assurance case reviewers are able to determine the level of trust of argument elements;
- Ability to express counter-arguments, so that the argumentation process becomes more comprehensible;
- Traceability from evidence to artifact, so that changes of system information/argumentation can be propagated throughout the assurance case, to enable incremental certification;
- Automated assurance case instantiation, to link system information with failure modes to create concrete assurance cases.

We also provided a definitive exposition of SACM to explain its intended usage via examples. SACM has been sufficiently explained in this paper although extensive examples cannot be fully provided.

We provided our version of GSN and CAE metamodels, which are compliant to SACM in the sense that users of these metamodels are able to exploit the facilities provided by SACM whilst still using GSN/CAE notations that they are familiar with. We also provide comprehensible model-to-model transformations from GSN/CAE to SACM to enable the interoperability from GSN/CAE to SACM.

We also briefly discussed the Assurance Case Modelling Environment (ACME) - a graphical modelling tool created based on our version of the GSN metamodel. With ACME, we explained what is supported when assurance cases become model-based, and what can be done in the future work of ACME. ACME acts as a transitional solution from conventional GSN diagram creation to SACM model-based system assurance. ACME provides easy-to-use support for SACM facilities, as well as automated model-to-model transformation from GSN to SACM.

SACM provides a solid foundation for model-based system assurance, due to the variety of features that have been evaluated and added to it, which are based on experiences from two well-established assurance case notations: GSN and CAE. Model-based assurance case is the key to assure Open Adaptive Systems (such as Cyber-Physical Systems), for it enables system assurance to be performed at runtime, which entails automated system assurance case integration, and automated reasoning of assurance cases.

**Acknowledgements** This work is supported by the European Union' s Horizon 2020 research and innovation programme through the DEIS project (grant agreement No 732242).